\documentclass[12pt]{article}

\RequirePackage{amsthm,amsmath,amsfonts,amssymb}
\RequirePackage[authoryear]{natbib}
\usepackage[usenames]{color}
\RequirePackage[colorlinks,linkcolor=blue,citecolor=blue,urlcolor=blue]{hyperref}
\usepackage[margin=1in]{geometry}
\usepackage{booktabs}
\usepackage{url}
\usepackage{graphicx} 
\usepackage{amsmath}
\usepackage{amsthm}
\usepackage{bbm}
\usepackage{stackengine}
\usepackage{amssymb}
\usepackage{mathtools}
\usepackage{setspace}
\usepackage{dsfont}
\usepackage{color,soul}
\usepackage{xcolor}
\usepackage{caption}
\usepackage{subcaption}
\usepackage[T1]{fontenc}   
\usepackage{multirow}
\usepackage{mathrsfs}
\usepackage{bm}
\usepackage{fancyhdr}
\usepackage{tabularx}
\usepackage{tabulary}
\usepackage{algorithm}
\usepackage{algpseudocode}
\usepackage{tikz}
\usetikzlibrary{positioning}
\usepackage{diagbox}
\usepackage{enumitem}
\usepackage{hyperref}
\usepackage{xr}
\usepackage{footmisc}
\usepackage[normalem]{ulem}
\usepackage{makecell}

\newcommand{\bM}{{\mathbf M}}

\newcommand{\bX}{{\mathbf X}}

\newcommand{\beq}{\begin{eqnarray*}}
\newcommand{\eeq}{\end{eqnarray*}}
\newcommand{\beqn}{\begin{eqnarray}}
\newcommand{\eeqn}{\end{eqnarray}}

\numberwithin{equation}{section}
\theoremstyle{plain}

\newtheorem{theorem}{Theorem}
\newtheorem{proposition}{Proposition}
\theoremstyle{remark}

\newtheorem{remark}{Remark}
\newtheorem{assumption}{Assumption}

\newcommand{\argmin}{\mathop{\text{\normalfont{argmin}}}}
\newcommand{\Cov}{\mathop{\text{\normalfont{Cov}}}}

\newcommand{\Var}{\mathop{\text{\normalfont{Var}}}}

\newcommand{\wh}{\widehat}

\title{Model-Free Inference for Characterizing Protein Mutations through a Coevolutionary Lens}

\author{
Fan Yang$^{1}$ \and
Zhao Ren$^{1}$\thanks{Corresponding author: zren@pitt.edu} \and
Wen Zhou$^{2}$ \and
Kejue Jia$^{3}$ \and
Robert Jernigan$^{4}$
}

\date{}

\begin{document}
\maketitle

\begin{center}
$^{1}$ Department of Statistics, University of Pittsburgh \\
$^{2}$ Department of Biostatistics, School of Global Public Health, New York University \\
$^{3}$ Department of Molecular, Cellular and Developmental Biology, Yale University \\
$^{4}$ Department of Biochemistry, Biophysics and Molecular Biology, Iowa State University  
\end{center}

\begin{abstract}
Multiple sequence alignment (MSA) data play a crucial role in the study of protein mutations, with contact prediction being a notable application. Existing methods are often model-based or algorithmic and typically do not incorporate statistical inference to quantify the uncertainty of the prediction outcomes. To address this, we propose a novel framework that transforms the task of contact prediction into a statistical testing problem. Our approach is motivated by the partial correlation for continuous random variables. With one-hot encoding of MSA data, we are able to construct a partial correlation graph for multivariate categorical variables. In this framework, two connected nodes in the graph indicate that the corresponding positions on the protein form a contact. A new spectrum-based test statistic is introduced to test whether two positions are partially correlated. Moreover, the new framework enables the identification of amino acid combinations that contribute to the correlation within the identified contacts, an important but largely unexplored aspect of protein mutations. Numerical experiments demonstrate that our proposed method is valid in terms of controlling Type I errors and powerful in general. Real data applications on various protein families further validate the practical utility of our approach in coevolution and mutation analysis.
\end{abstract}

\noindent%

\vfill

\newpage



\section{Introduction}
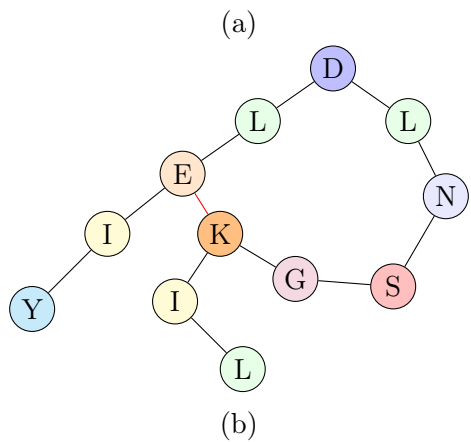
\begin{figure}[!h]
    \centering
\begin{subfigure}{0.6\textwidth}
\begin{tikzpicture}
    \node at (0,0) {\scriptsize\texttt{\hspace{-3.9cm}{\color{blue} 23}\hspace{1.8cm}{\color{blue}30}}};
    \node at (0,-0.3) {\scriptsize\textcolor{red}{\texttt{Y I E L D L N S G K I L E S F R P E E R F P M M S}}};
    \node at (0,-0.6) {\scriptsize\texttt{Y A V L Q F D D E E E I A S K G E A T V H S S A S}};
    \node at (0,-0.9) {\scriptsize\texttt{V A I K D L S G S K E L - H L G S R Q P Y M P A S}};
     \node at (0,-1.2) {\scriptsize\texttt{V D I K D L A G G A E V - L L G S R Q S Y M P A S}};
      \node at (0,-1.5) {\scriptsize\texttt{V Y A V D T A S G R S V - E H R P N E R F P F C S}};
      \node at (0,-1.8) {\scriptsize\texttt{V A V L D T G T G Q R F - G H R A D E R F P M C S}};
      \node at (0,-2.1) {\scriptsize\texttt{V A I I D S Q S G R Q W - L H R A D E R F P L C S}};
      \node at (0,-2.4) {\scriptsize\texttt{V A L I N T A D N S Q I - L Y R G D E R F A M C S}};
    \node at (0,0) {\scriptsize\texttt{\hspace{-0.0cm}21\hspace{6.9cm}45}};     
\end{tikzpicture}
\caption{}
\end{subfigure}%

\begin{subfigure}{0.4\textwidth}
\centering
\begin{tikzpicture}
    \node[circle, draw,fill=cyan!20,font=\small, minimum size=17pt, inner sep=0pt] (Y) at (0,0) {Y};
    \node[circle, draw,fill=yellow!20,font=\small, minimum size=17pt, inner sep=0pt] (I) at (1,1) {I};
    \node[circle, draw,fill=orange!20,font=\small, minimum size=17pt, inner sep=0pt] (E) at (2,1.8) {E};
    \node[circle, draw,fill=green!10,font=\small, minimum size=17pt, inner sep=0pt] (L) at (3,2.5) {L};
    \node[circle, draw,fill=blue!25,font=\small, minimum size=17pt, inner sep=0pt] (D) at (4,3.2) {D};
    \node[circle, draw,fill=green!10,font=\small, minimum size=17pt, inner sep=0pt] (L2) at (5,2.5) {L};
    \node[circle, draw,fill=blue!9,font=\small, minimum size=17pt, inner sep=0pt] (N) at (5.5,1.5) {N};
    \node[circle, draw,fill=pink,font=\small, minimum size=17pt, inner sep=0pt] (S) at (4.8,0.3) {S};
    \node[circle, draw,fill=purple!15,font=\small, minimum size=17pt, inner sep=0pt] (G) at (3.5,0.4) {G};
    \node[circle, draw,fill=orange!50,font=\small, minimum size=17pt, inner sep=0pt] (K) at (2.5,1) {K};
    \node[circle, draw,fill=yellow!20,font=\small, minimum size=17pt, inner sep=0pt] (I2) at (1.9,0.1) {I};
    \node[circle, draw,fill=green!10,font=\small, minimum size=17pt, inner sep=0pt] (L3) at (2.8,-0.8) {L};

    \draw (Y) -- (I);
    \draw (I) -- (E);
    \draw (E) -- (L);
    \draw (L) -- (D);
    \draw (D) -- (L2);
    \draw (L2) -- (N);
    \draw (N) -- (S);
    \draw (S) -- (G);
    \draw (G) -- (K);
    \draw (K) -- (I2);
    \draw (I2) -- (L3);
    \draw[red] (E) -- (K);
\end{tikzpicture}
\caption{}
\end{subfigure}

\caption{(a) Protein sequences from the class A beta-lactamase family, with positions $21$-$45$ displaye. Positions $23$ and $30$ are spatially close and known to contact structurally. (b) Illustration of the Beta-lactamase TEM sequence (sequence in red in Plot (a), positions $21$-$32$), featuring Glutamic acid (E) at position $23$ and Lysine (K) at position $30$. The corresponding crystallization structure appears in Figure \ref{fig:heatmap}. }
\label{fig::msa}
\end{figure}

In protein structures, residues in close contact tend to ``coevolve'', meaning that when one residue mutates, its
interacting partners tend to undergo compensatory mutations to preserve structural or functional stability \citep{jia2021new,storz2018compensatory}. Intra-molecule coevolution can be reflected by mutation dependencies among residues within a protein. It has been successfully utilized to predict protein residue contacts, amino acid residues that interact in close spatial proximity (Panel (b) in Figure \ref{fig::msa}), which is crucial for predicting three-dimensional protein structures, protein-protein interactions, and protein dynamics \citep{marks2012protein,jia2023functional}. In addition, quantifying coevolution among residues provides important insights into how mutations at specific positions influence the overall function of the protein under evolutionary selection \citep{hopf2017mutation}.

Coevolution methods often use multiple sequence alignment (MSA) as the input data, which contains aligned residue positions across evolutionarily related protein sequences. For example, Panel (a) of Figure \ref{fig::msa} shows a portion of the MSA data for the class A beta-lactamase family. In the alignment, each row represents a protein sequence, with twenty distinct letters\footnote{\bf A, C, D, E, F, G, H, I, K, L, M, N, P, Q, R, S, T, V, W, Y} encoding specific amino acids and dashed lines indicating alignment gaps. 

A key challenge in coevolution analysis is how to remove the so-called ``indirect coupling'',  where the influence of changes in amino acids at one position can propagate through the surrounding neighborhoods, reaching positions not only in close contact by also spatially farther from the original mutation site. In contrast to indirect coupling, direct coupling refers to residue pairs that have immediate interaction due to mutation and are often spatially close, with the influence of other residues being well accounted for. Therefore, direct coupling information can be effectively used to infer residue contacts.

Many widely used sequence-based approaches for contact prediction, such as correlated mutation analysis (CMA) and mutual information (MI) \citep{dunn2008mutual,ashkenazy2010reducing}, fail to disentangle direct coupling effects from indirect ones, potentially leading to false positives in predictions. Direct Coupling Analysis (DCA, \cite{morcos2011direct}) was introduced to address this issue by using the Potts model to represent the distribution of protein sequences. However, as a model-based approach, it is sensitive to model misspecification, which can lead to inaccurate residue contact predictions, even under mild model misspecifications. Moreover, parameter inference under the Potts model can be challenging, and to the best of our knowledge, no rigorous inferential framework has been established for the Potts model. In contrast, partial correlation \citep{fisher1924distribution} is a model-free measure that assesses the association between two continuous random variables while controlling for the influence of others. Conceptually, it is similar to the process of eliminating indirect coupling effects, making it a natural choice for contact predictions. This motivates us to develop a properly defined partial correlation for multivariate categorical variables as a measure of direct coupling strength in protein sequences. As shown in Section \ref{sec::contact prediction}, built up on a novel partial correlation for MSA data, our procedure for quantifying the direct coupling effect improves the accuracy of residue contact predictions at the position level. In Section \ref{sec::amcidcomp}, we further demonstrate that our approach can explain compensatory mutations at the amino acid level and identify specific amino acid combinations contributing to coevolved residues.

In addition, quantifying direct coupling among residues can assist in predicting mutation effects. Traditionally, amino acid conservation at a single position is used to infer the likelihood of a mutation being deleterious. However, a harmful mutation at one position can be compensated by mutations at other positions, and conversely, a seemingly neutral mutation may become more harmful due to interdependencies between positions \citep{jia2021new}. Therefore, the amino acid composition across multiple coevolved positions defines a more general conservation of protein sequence \citep{storz2018compensatory}. In Section \ref{sec::MutPred}, we show that incorporating quantified direct coupling information at the amino acid level---capturing dependencies between different amino acid types across coevolved positions, which is under explored by existing approaches---provides new insights into the mechanism of compensatory mutations and enhances mutation effect assessment.

In the following subsections, we will formally introduce how partial correlation can be constructed in the context of MSA data. We refer to our method as a model-free approach, since it neither imposes any distributional assumptions on protein sequences nor focuses on the estimation and inference of model parameters. Rather, we employ the idea of partial correlation, which is well defined under the existence of second moments and a positive-definite precision matrix. These are considerably weaker assumptions.

\subsection{Data format and related notations}\label{sec1.1}
Partial correlation is not directly applicable to categorical variables, such as those in MSA data. Here, we first transform the MSA data before we formally define the partial correlation for categorical variables in Section \ref{sec12}. Suppose there are $N$ sequences and $m$ positions. Each position $i$ has $d_i$ unique amino acids. It is expanded into an $N\times d_i$ matrix, where each column represents one amino acid, and the entry is $1$ if the corresponding sequence has that amino acid at position $i$, otherwise, $0$. Note that the column representing gaps is dropped to prevent identifiability issues. The resulting $m$ matrices are concatenated to form the final data set $\mathbf{X}$ with $D$ columns, where $D\coloneqq d_1+d_2+\cdots+d_m$.  Each column of $\mathbf{X}$ is then standardized to have a mean of $0$ and a unit variance. A diagram illustrating the one-hot encoding procedure is provided in the supplementary material.

Let $\mathcal{G}_i$ represent the set of indices for the columns in $\mathbf{X}$ associated with position $i$. Then, we divide $\mathbf{X}$ into $m$ submatrices as $\mathbf{X}=(\mathbf{X}_{\boldsymbol{\cdot},\mathcal{G}_1},\dots,\mathbf{X}_{\boldsymbol{\cdot},\mathcal{G}_m})$. Additionally, define $\mathcal{G}_{i,j}=\mathcal{G}_i\cup \mathcal{G}_j$, $\mathcal{G}^c_{i,j}=\{1,\dots,D\}\setminus\mathcal{G}_{i,j}$ and $\mathcal{G}^c_{i}=\{1,\dots,D\}\setminus\mathcal{G}_{i}$.

\subsection{Partial correlations for multivariate categorical variable}
\label{sec12}

In the context of a graph, partial correlation indicates whether two nodes are connected after accounting for the linear effects of other nodes. For example, for random variables ${X_1,\ldots, X_6}$ modeled as nodes in Figure 2(a), two nodes are connected if and only if their partial correlation is nonzero. When all $X_i$ are normally distributed, the graph defined by partial correlations is a Gaussian graph, where edges represent conditional dependencies between nodes. A rich literature exists on the estimation and entry-wise inference of partial correlations or precision matrices, particularly in high-dimensional settings; however, those primarily focus on continuous random variables \citep{meinshausen2006high,deng2009large,ren2015asymptotic,jankova2017honest,ren2022gaussian,zhang2023high,lee2023conditional}, and their direct application to one-hot transformed data can lead to misleading interpretations (e.g., PSICOV \citep{jones2012psicov}, as discussed later in this section.

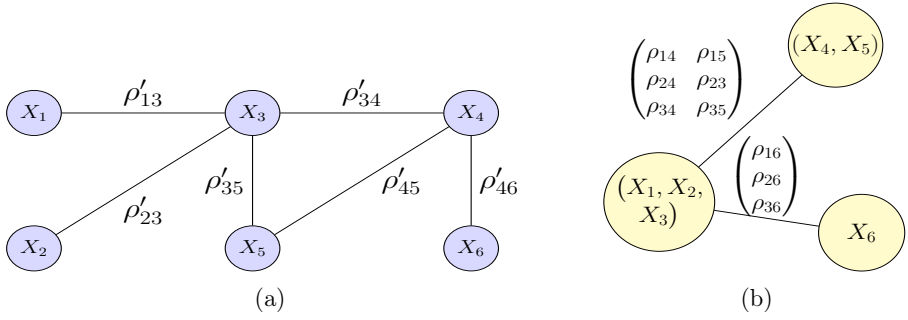
\begin{figure}[h]
    \centering
     \scalebox{.85} 
 {
    \begin{subfigure}[t]{.5\textwidth}
        \centering
         \resizebox{\linewidth}{3.1cm}{%
        \begin{tikzpicture}[node distance=2cm and 3cm]
    \node[circle, fill=blue!15, draw] (X1) {$X_1$};
    \node[circle, draw, fill=blue!15, right=of X1] (X3) {$X_3$};
    \node[circle, draw, fill=blue!15, right=of X3] (X4) {$X_4$};
    \node[circle, draw, fill=blue!15, below=of X1] (X2) {$X_2$};
    \node[circle, draw, fill=blue!15, below=of X3] (X5) {$X_5$};
    \node[circle, draw, fill=blue!15, below=of X4] (X6) {$X_6$};
    
    \draw (X1) -- node[above, midway] {\Large $\rho'_{13}$} (X3);
    \draw (X2) -- node[below, midway, inner sep=10pt] {\Large $\rho'_{23}$} (X3);
    \draw (X3) -- node[above, midway] {\Large $\rho'_{34}$} (X4);
    \draw (X3) -- node[left, midway] {\Large $\rho'_{35}$} (X5);
    \draw (X4) -- node[right, midway, inner sep=10pt] {\Large $\rho'_{45}$} (X5);
    \draw (X4) -- node[right, midway] {\Large $\rho'_{46}$} (X6);
\end{tikzpicture}
}
        \caption{ }
    \end{subfigure}%
    \hspace{0.95cm}
    \begin{subfigure}[t]{.29\textwidth}
        \centering
         \resizebox{\linewidth}{4.2cm}{%
        \begin{tikzpicture}[
    every node/.style={ inner sep=1pt, minimum size=2.5cm, align=center},
    every edge/.style={draw, -},
    label distance=2mm,
    node distance=2cm and 3cm 
]

    \node[circle, fill=yellow!25, draw] (X123) at (0,0) {\LARGE $\big(X_1, X_2,$\\ \LARGE $X_3\big)$};
    \node[circle, fill=yellow!25, draw, above right=of X123, yshift=1cm] (X45) {\LARGE$\left(X_4, X_5\right)$};  
    \node[circle, draw, fill=yellow!25, right=of X123, yshift=-1cm] (X6) {\LARGE $X_6$};  

    \draw (X123) -- (X45) node[midway,  above left] {\LARGE $\left(\begin{matrix} \rho_{14} &~ \rho_{15} \\ \rho_{24} &~ \rho_{23} \\ \rho_{34} &~ \rho_{35} \end{matrix}\right)$};
    \draw (X123) -- (X6) node[midway, above] {\LARGE $\left(\begin{matrix} \rho_{16} \\ \rho_{26} \\ \rho_{36} \end{matrix}\right)$};

\end{tikzpicture}
}
        \caption{ }
    \end{subfigure}
    }
\caption{Demonstration of two different partial correlations.} 
\label{graph-example}
\end{figure}

Adopting partial correlation to study protein coevolution,  particularly to account for indirect coupling effects, presents an unprecedented challenge: with the transformed data $\mathbf{X}$, the partial correlation between two positions (nodes in a graph) is no longer a scalar but a matrix, as each position is represented by a random vector
(see Figure \ref{graph-example}(b)). This differs fundamentally from the conventional scalar partial correlation shown in Figure \ref{graph-example}(a). For example, in Panel (a), the partial correlation $\rho'_{35}$ between $X_3$ and $X_5$ accounts for the linear effects of $X_1, X_2, X_4,$ and $X_6$. In contrast, in Panel (b), the entry $\rho_{35}$ in the partial correlation matrix, formally defined below, between $(X_1, X_2, X_3)$ and $(X_4, X_5)$ only controls for the effect of $X_6$. The partial correlation graph in Panel (b) is more biologically meaningful in the context of contact prediction. This is because, when calculating the partial correlation between two amino acids at two positions, we aim to avoid adjusting for the effects of other types of amino acids at those same positions.

Suppose each row of $\mathbf{X}$ is independent, following the same distribution of $\mathbf{Z}\in \mathbb{R}^D$ with a positive definite precision matrix, $\mathbf{\Omega}=(\Cov(\mathbf{Z}))^{-1}$. For a given pair $(i,j)$, we take a local regression approach to regress $\mathbf{Z}_{\mathcal{G}_{i,j}}$ against $\mathbf{Z}_{\mathcal{G}^c_{i,j}}$. It holds that
$$
\mathbf{Z}_{\mathcal{G}_{i,j}}
=-\left[\mathbf{\Omega}_{\mathcal{G}_{i,j},\mathcal{G}_{i,j}}\right]^{-1}\mathbf{\Omega}_{\mathcal{G}_{i,j},\mathcal{G}_{i,j}^c}\mathbf{Z}_{\mathcal{G}_{i,j}^c}+\boldsymbol{\mathcal{E}}_{(i,j)}$$ \citep{izenman2008modern}, 
where the oracle error $\boldsymbol{\mathcal{E}}_{(i,j)}$ is uncorrelated to $\mathbf{Z}_{\mathcal{G}^c_{i,j}}$. Furthermore, denote by $\boldsymbol{\mathcal{E}}_{(i,\backslash(i,j))}$ and  $\boldsymbol{\mathcal{E}}_{(j,\backslash(i,j))}$ the errors of regressing $\mathbf{Z}_{\mathcal{G}_i}$ and $\mathbf{Z}_{\mathcal{G}_j}$ against $\mathbf{Z}_{\mathcal{G}^c_{i,j}}$, respectively; i.e., $\boldsymbol{\mathcal{E}}_{(i,j)}=(\boldsymbol{\mathcal{E}}^T_{(i,\backslash(i,j))},\boldsymbol{\mathcal{E}}^T_{(j,\backslash(i,j))})^T$. Let $\operatorname{diag}(\bM)$ be the diagonal matrix formed from the diagonal entries of a square matrix $\bM$. Then, the partial correlation between $\mathbf{Z}_{\mathcal{G}_i}$ and $\mathbf{Z}_{\mathcal{G}_j}$ is defined as $$\mathbf{P}^{(i,j)}:=[\operatorname{diag}\left(\Cov(\boldsymbol{\mathcal{E}}_{(i,\backslash(i,j))})\right)]^{-1/2}\Cov(\boldsymbol{\mathcal{E}}_{(i,\backslash(i,j))},\boldsymbol{\mathcal{E}}_{(j,\backslash(i,j))})[\operatorname{diag}\left(\Cov(\boldsymbol{\mathcal{E}}_{(j,\backslash(i,j))})\right)]^{-1/2}.$$  Referring to Panel (b) of Figure \ref{graph-example}, it represents three positions with one, two, and three different amino acids, respectively. For example, when calculating $\rho_{14}$, we regress $(X_1, X_2, X_3)$ and $(X_4, X_5)$ separately against $X_6$. However, we do not regress out the effects of $(X_2, X_3)$ and $X_5$, as they belong to the same nodes as $X_1$ and $X_4$, respectively. The partial correlation is then computed from the regression residuals, as described earlier. It is worth noting that the regression-based definition of partial correlation adopted here applies beyond the Gaussian setting and is well-defined whenever second moments exist. In general distributions, partial correlation continues to quantify linear conditional association, although it does not fully characterize conditional independence.

One commonly used method, PSICOV \citep{jones2012psicov}, can be related to the partial correlation defined for categorical variables above. It employs the precision matrix, estimated by graphical Lasso \citep{friedman2008sparse}, instead of the proposed partial correlation in our approach. With the $\wh{\mathbf{\Omega}}=(\wh{\omega}_{t_1,t_2})_{t_1,t_2=1}^{D}$ estimated from MSA sequences (rows of $\mathbf{X}$), it introduces score $T_{ij}^{\mathrm{contact}}=\sum_{t_1\in \mathcal{G}_i, t_2\in \mathcal{G}_j} |\wh{\omega}_{t_1,t_2}|$ to measure the direct coupling effect between positions $i,j$. It is well known that the estimated partial correlation between any two columns of $\bX$, $\mathbf{X}_{\boldsymbol{\cdot},t_1}$ and $\mathbf{X}_{\boldsymbol{\cdot},t_2}$, admits $\wh{\rho}_{t_1,t_2}=-\wh{\omega}_{t_1,t_2}/\sqrt{\wh{\omega}_{t_1,t_1}\wh{\omega}_{t_2,t_2}}$. Thus, the PSICOV score somehow measures the strength of partial correlations but does not standardize appropriately. Also, the PSICOV score can be misleading as it is based on the model in Figure \ref{graph-example}(a). That is, for any $t_1\in\mathcal{G}_i$ and $t_2\in\mathcal{G}_j$ for positions $i$ and $j$, $\wh{\omega}_{t_1,t_2}$ does not only adjust for the effects from positions other than $i,j$, but also adjusts for the effects from other types of amino acids on positions $i$ and $j$, which however is biologically relevant.

\subsection{Statistical test for protein coevolutions}\label{formtest}

We characterize protein coevolutions by designing a novel statistical test on partial correlation $\mathbf{P}^{(i,j)}$. Detecting a direct coupling effect between positions $i$ and $j$ is transformed into testing if the partial correlation between $\mathbf{Z}_{\mathcal{G}_i}$ and $\mathbf{Z}_{\mathcal{G}_j}$ is zero. Equivalently, it can be formally stated as the null hypothesis, $H_0: \mathbf{P}^{(i,j)}=0$. We will construct a spectrum-based test statistic using the fitted residuals to tackle this problem and provide theoretical guarantees for the proposed test.

Additionally, our framework allows for a closer examination of the contribution of amino acid combinations to contacts and mutations. By computing their estimated partial correlations using fitted residuals, we can infer whether specific amino acid combinations across positions $(i,j)$ are significantly partially correlated. This can help decipher compensatory mutations within contacts, which enhance our understanding of the interaction of contact residues at the amino acid level.

Although one-hot encoding transforms each categorical position into a block of binary variables, this does not turn the problem into a standard continuous-data setting. Existing methods that operate on one-hot encoded MSA data (e.g., PSICOV) implicitly rely on partial correlations but do not account for the block structure inherent in categorical variables, leading to interpretations that can be misleading.

Our contribution is to formally define partial correlation for multivariate categorical variables in a way that preserves this block structure and yields an interpretable notion of conditional dependence between positions. Once this definition is established, the subsequent regression-based steps coincide with those used for continuous variables, but the interpretation is fundamentally different.

Overall, our contributions are threefold: (1) To the best of our knowledge, we are the first to introduce partial correlation as a method for quantifying the strength of direct coupling and apply it to contact prediction. Unlike existing methods, our approach is inference-based. (2) Using one-hot encoded MSA data, we propose a framework to reconstruct a partial correlation graph, where each edge is represented by a matrix, with each entry corresponding to the partial correlation between amino acids at two positions. (3) Finally, our framework not only leads to an inference-based method that improves contact prediction but also identifies amino acid combinations contributing to the contacts and generates novel features that enhance mutation effect prediction.

\subsection{Organization and more notations.} In Section \ref{sec:test}, we introduce the estimation procedure for obtaining the fitted residuals, followed by the testing framework for residue-residue contact prediction and the identification of highly correlated amino acid combinations. Section \ref{sec3} describes more details on the implementations. Section \ref{sec:simulation} presents simulation results to validate the efficacy of our spectrum-based test statistics. In Section \ref{sec:real}, we apply our methodology to real data, and focus on contact predictions and the identification of contributing amino acid combinations. Additionally, we briefly discuss how the estimated partial correlations can be used to generate new features, which improve the performance of Evolutionary Scale Modeling (ESM) for mutation effect prediction \citep{rives2021biological}, as shown by results in Section \ref{sec::MutPred}. 

\section{Statistical framework for inferring protein coevolution}\label{sec:test} 

\subsection{Estimation procedure} \label{sec: estimation procedure}

Consider the following regression equations,
\begin{equation}\label{2.1reg}
    \begin{split}
&\mathbf{X}_{\boldsymbol{\cdot},\mathcal{G}_i}=\mathbf{X}_{\boldsymbol{\cdot},\mathcal{G}^c_{i,j}}\mathbf{B}_{(i,\backslash\{i,j\})}+\mathbf{E}_{(i,\backslash\{i,j\})}\\
&\mathbf{X}_{\boldsymbol{\cdot},\mathcal{G}_j}=\mathbf{X}_{\boldsymbol{\cdot},\mathcal{G}^c_{i,j}}\mathbf{B}_{(j,\backslash\{i,j\})}+\mathbf{E}_{(j,\backslash\{i,j\})}.
    \end{split}
\end{equation}
To illustrate our estimation procedure for Equation \eqref{2.1reg} across all positions, we focus on the regression of $\mathbf{X}_{\boldsymbol{\cdot},\mathcal{G}_i}$ against $\mathbf{X}_{\boldsymbol{\cdot},\mathcal{G}^c_{i,j}}$ for a specific pair of positions, $(i,j)$. For simplicity, we refer to the $(D-d_i-d_j)\times d_i$ coefficient matrix $\mathbf{B}_{(i,\backslash\{i,j\})}$ as $\mathbf{B}$. The rows of $\mathbf{B}$ corresponding to the same position naturally form a group. Specifically, let $\mathbf{B}_{g_{i'}}$ denote the $d_{i'} \times d_i$ submatrix within $\mathbf{B}$, which contains the coefficients associated with position $i'$. Thus, $\mathbf{B}$ consists of $(m-2)$ such submatrices.

Most protein families of interest have hundreds of positions in their sequences \citep{finn2014pfam,el2019pfam}, which makes our task a genuinely high-dimensional problem entangled with inherent group structures. On the other hand, the mutual influence between most pairs of positions is limited, primarily due to their substantial spatial separation. As a result, the partial correlation matrices $\mathbf{P}^{(i',j')}\text{'s}$ are expected to be zero for the majority of these pairs, which  implies group sparsity within $\mathbf{B}$ in \eqref{2.1reg}. As noted in Section \ref{sec12}, we have $[\mathbf{B}_{(i,\backslash\{i,j\})},\mathbf{B}_{(j,\backslash\{i,j\})}]=-\mathbf{\Omega}_{\mathcal{G}^c_{i,j},\mathcal{G}_{i,j}}[\mathbf{\Omega}_{\mathcal{G}_{i,j},\mathcal{G}_{i,j}}]^{-1}$. Furthermore, it is known that $\Cov(\mathbf{\mathcal{E}}_{(i,j)})=[\mathbf{\Omega}_{\mathcal{G}_{i,j},\mathcal{G}_{i,j}}]^{-1}$ \citep[Section 6.3]{izenman2008modern}. Hence, $\mathbf{P}^{(i,j)}=0$ is equivalent to $\mathbf{\Omega}_{\mathcal{G}_i,\mathcal{G}_j}=0$. We define the maximum node degree as the largest number of non-zero $\mathbf{P}^{(i',j')}$, $j'=\{1,\dots,m\}\setminus\{i'\}$ across all positions $i'\in \{1,\dots,m\}$. From the earlier observation, if the maximum node degree is no greater than $s$, then there are at most $2s$ non-zero submatrices $\mathbf{B}_{g_{i'}}\text{'s}$ in the coefficient matrix $\mathbf{B}$.

In light of the sparsity in the coefficient matrix, we adopt the multivariate group Lasso \citep{li2015multivariate} to estimate $\mathbf{B}_{(i,\backslash\{i,j\})}$. Henceforth, let $\mathcal{P}_{i,j}^c=\{1,\dots,m\}\setminus\{i,j\}$ be the index set excluding positions $i$ and $j$. With $\|\bM\|_2 = \sum_{t_1,t_2} M_{t_1,t_2}^2$ for matrix $\bM$, estimator $\wh{\mathbf{B}}_{(i,\backslash\{i,j\})}$ is defined as follows. \begin{equation}\label{objective}
	  \wh{\mathbf{B}}_{(i,\backslash\{i,j\})}=\underset{\breve{\mathbf{B}}\in \mathbb{R}^{(D-d_i-d_j)\times d_i}}{\argmin} \frac{1}{2N}\|\mathbf{X}_{\boldsymbol{\cdot},\mathcal{G}_i}-\mathbf{X}_{\boldsymbol{\cdot},\mathcal{G}^c_{i,j}}\breve{\mathbf{B}}\|_2^2+ \sum_{i'\in \mathcal{P}_{i,j}^c} \lambda_{i'}\|\breve{\mathbf{B}}_{g_{i'}}\|_2. 
\end{equation}	
Similarly, we can regress $\mathbf{X}_{\boldsymbol{\cdot},\mathcal{G}_j}$ against $\mathbf{X}_{\boldsymbol{\cdot},\mathcal{G}^c_{i,j}}$ to obtain $\wh{\mathbf{B}}_{(j,\backslash\{i,j\})}$. It is worth noting that a more general objective function may include both group penalty and individual penalty, which can yield a more interpretable estimator by allowing only a subset (rather than all) of amino acids at two positions to be associated. However, we did not adopt this formulation because preliminary experiments indicated that incorporating an individual penalty does not necessarily improve prediction performance and introduces additional tuning parameters. Moreover, since the group size in our application is fixed, the estimation error bound stated in Theorem F.1 of the supplementary material (with only the group penalty) matches the order of the bound in Theorem 2 of \cite{li2015multivariate} (which includes both penalties). Consequently, this choice does not affect any downstream theoretical analysis.

In high-dimensional linear regression, it is well-established that coefficient estimation tends to be less accurate and converges more slowly compared to the parametric rate observed in low-dimensional settings. Consequently, without additional processing, the estimated coefficients cannot be directly used for statistical inferences. For an illustrative example, see \cite{zhang2014confidence}. However, prediction is generally easier than coefficient estimation \citep{ren2015asymptotic}. Given the estimated coefficient matrix $\wh{\mathbf{B}}_{(i,\backslash\{i,j\})}$, the fitted residual matrix is computed as $\wh{\mathbf{E}}_{(i,\backslash\{i,j\})}=\mathbf{X}_{\boldsymbol{\cdot},\mathcal{G}_i}-\mathbf{X}_{\boldsymbol{\cdot},\mathcal{G}^c_{i,j}}\wh{\mathbf{B}}_{(i,\backslash\{i,j\})}$. We can compute $\wh{\mathbf{E}}_{(j,\backslash\{i,j\})}$ likewise. The residual matrices can then be used to construct our test statistic. Under certain mild assumptions on the sparsity of the precision matrix $\mathbf{\Omega}$, we will justify in Theorem \ref{thm1} that the test statistic derived from fitted residuals is valid. It's worth noting that there are alternative regression procedures in the literature. For example, \cite{xia2018multiple} study testing problems for submatrices of Gaussian precision matrices by performing separate regressions of each variable against all others (including those within the same block), without leveraging the block structure. Their approach relies on residuals to construct $\ell_2$-norm–based statistics but requires an additional de-biasing step. In contrast, we adopt multivariate regression with a group penalty, which avoids the need for de-biasing and makes the subsequent construction of test statistics from fitted residuals more straightforward.

\subsection{Testing framework for contact prediction}

\subsubsection{Background on the CCA-related test}\label{cca}
  
Without loss of generality, we focus on the pair of positions $(i,j)=(1,2)$ and assume that $d_1\leq d_2$. In addition, we introduce the following notations: $\Cov(\boldsymbol{\mathcal{E}}_{(1,\backslash(1,2))})=\breve{\mathbf{\Sigma}}_{11}$, $\Cov(\boldsymbol{\mathcal{E}}_{(2,\backslash(1,2))})=\breve{\mathbf{\Sigma}}_{22}$, $\Cov(\boldsymbol{\mathcal{E}}_{(1,\backslash(1,2))},\boldsymbol{\mathcal{E}}_{(2,\backslash(1,2))})=\breve{\mathbf{\Sigma}}_{12}$ and $\breve{\mathbf{\Sigma}}_{21}=(\breve{\mathbf{\Sigma}}_{12})^T$. Testing whether $\mathbf{P}^{(1,2)}=0$ is equivalent to testing the null hypothesis  $\eta_1^2=\dots=\eta_{d_1}^2=0$, where $1>\eta^2_1\geq \dots \geq \eta^2_{d_1}\geq 0$ are the eigenvalues of the matrix $(\breve{\mathbf{\Sigma}}_{11})^{-1}\breve{\mathbf{\Sigma}}_{12}(\breve{\mathbf{\Sigma}}_{22})^{-1}\breve{\mathbf{\Sigma}}_{21}$. The square roots, $\eta_{\ell}\text{'s}$, are called the population canonical correlation coefficients. We denote $\mathbf{S}_{11}$, $\mathbf{S}_{22}$, and $\mathbf{S}_{21}$ as the empirical counterparts of $\breve{\mathbf{\Sigma}}_{11}$, $\breve{\mathbf{\Sigma}}_{22}$, and $\breve{\mathbf{\Sigma}}_{21}$, respectively. They are computed based on the oracle residuals; that is,
\begin{equation}
    \mathbf{S}=\frac{1}{N}(\mathbf{E}_{(1,\backslash\{1,2\})},\mathbf{E}_{(2,\backslash\{1,2\})})^T(\mathbf{E}_{(1,\backslash\{1,2\})},\mathbf{E}_{(2,\backslash\{1,2\})})=\begin{pmatrix}
	\mathbf{S}_{11} & \mathbf{S}_{12} \\
	\mathbf{S}_{21} & \mathbf{S}_{22}
 \end{pmatrix}. \label{matrix_s}
\end{equation}

Subsequently, we use $1>\wh{\eta}^2_1\geq \dots \geq \wh{\eta}_{d_1}^2\geq 0$ to denote the eigenvalues of the matrix $(\mathbf{S}_{11})^{-1}\mathbf{S}_{12}(\mathbf{S}_{22})^{-1}\mathbf{S}_{21}$. The test statistic, Wilk's lambda \citep{anderson1958introduction}, is then defined as $T^*_{1,2}=-N\log\prod_{\ell=1}^{d_1}(1-\wh{\eta}^2_{\ell})$. \cite{muirhead1980asymptotic} investigated the limiting  distribution of $T^*_{1,2}$ under the assumption of finite fourth moments of $\boldsymbol{\mathcal{E}}_{(1,2)}$. Under the null hypothesis $\eta_1^2=\dots=\eta_{d_1}^2=0$, it is shown that $T^*_{1,2}=\sum_{t_1=1}^{d_1}\sum_{t_2=1}^{d_2} Q_{t_1,d_1+t_2}^2+\mathcal{O}_p(1/N^{\frac{1}{2}})$, where $Q_{t_1,d_1+t_2}=\sqrt{N}\times \left([\breve{\mathbf{\Sigma}}_{11}]^{-1/2}\mathbf{S}_{12}[\breve{\mathbf{\Sigma}}_{22}]^{-1/2}\right)_{t_1,t_2}$. We stack all  $Q_{t_1,d_1+t_2}\text{'s}$ into a $d_1d_2$-dimensional vector, defined as $$\mathbf{Q}=(Q_{1,d_1+1},\dots,Q_{1,d_1+d_2},Q_{2,d_1+1},\dots,Q_{2,d_1+d_2},\dots,Q_{d_1,d_1+1},\dots,Q_{d_1,d_1+d_2})^T.$$ General results for the oracle residual-based test statistic is given in Proposition \ref{prop1} below.
\begin{proposition}[Test statistic based on the oracle errors, \cite{muirhead1980asymptotic}]
Under the null hypothesis $H_0: \mathbf{P}^{(1,2)}=0$, we have that $$T^{*}_{1,2}=-N\log\prod_{\ell=1}^{d_1}(1-\wh{\eta}^2_{\ell})\overset{\mathcal{L}}{\to} \sum_{\ell'=1}^{d_1d_2} \Lambda_{\ell'} W_{\ell'} ^2,$$ where $W_{\ell'}\text{'s}\sim N(0,1)$ are independent, and $\Lambda_{\ell'}\text{'s}$ are the eigenvalues of $\Cov(\mathbf{Q})$ for $\ell'=1,\dots,d_1d_2$. \label{prop1}
\end{proposition} 

Note that in general, the limiting distribution of $T^*_{1,2}$ follows a weighted Chi-squared distribution. In the special case where $\boldsymbol{\mathcal{E}}_{(1,2)}$ is normally distributed, $\Cov(\mathbf{Q})$ reduces to the identity matrix under the null hypothesis, and the limiting distribution simplifies to a standard Chi-squared distribution.

\subsubsection{Main results on the test statistic} \label{sec:main thm}

For the problem of contact prediction and mutation analysis, we only have the fitted residuals, $\wh{\mathbf{E}}_{(1,\backslash\{1,2\})}$ and $\wh{\mathbf{E}}_{(2,\backslash\{1,2\})}$. Let $\wh{\mathbf{S}}$ denote the sample covariance matrix calculated based on $(\wh{\mathbf{E}}_{(1,\backslash\{1,2\})},\wh{\mathbf{E}}_{(2,\backslash\{1,2\})})$. We partition $\wh{\mathbf{S}}$ into four blocks, $\wh{\mathbf{S}}_{11},\wh{\mathbf{S}}_{12},\wh{\mathbf{S}}_{21}$, and $\wh{\mathbf{S}}_{22}$, and denote by $\wh{r}^2_1\geq \dots \geq\wh{r}^2_{d_1}$ the eigenvalues of $\wh{\mathbf{S}}^{-1}_{11}\wh{\mathbf{S}}_{12}\wh{\mathbf{S}}_{22}^{-1}\wh{\mathbf{S}}_{21}$. The test statistic, $T_{1,2}$, is then defined as $T_{1,2}=-N\log\prod_{\ell=1}^{d_1}(1-\wh{r}^2_{\ell})$.

To estimate $\Cov(\mathbf{Q})$ in Proposition \ref{prop1}, we use quantities calculated based on the residuals and define, for $k=1,\dots,N$, $$\wh{\mathbf{Y}}_k=(\wh{Y}_{k,1},\dots,\wh{Y}_{k,d_1+d_2})^T=\left(\left[\wh{\mathbf{E}}_{(1,\backslash\{1,2\})}\right]_{k,\boldsymbol{\cdot}}\wh{\mathbf{S}}_{11}^{-1/2},\left[\wh{\mathbf{E}}_{(2,\backslash\{1,2\})}\right]_{k,\boldsymbol{\cdot}}\wh{\mathbf{S}}_{22}^{-1/2}\right)^T.$$ Then $\Cov(Q_{t_1,d_1+t_2},Q_{t'_1,d_1+t'_2})$ can be estimated by $N^{-1}\sum_{k=1}^N\wh{Y}_{k,t_1}\wh{Y}_{k,d_1+t_2}\wh{Y}_{k,t'_1}\wh{Y}_{k,d_1+t'_2}$ (a detailed explanation of why this can be used as the estimator is provided in the proof of Theorem \ref{thm1} in the supplementary material). Using this, we obtain an estimator of $\Cov(\mathbf{Q})$, denoted as $\wh{\Cov(\mathbf{Q})}$. With the mild assumptions listed below, Theorem \ref{thm1} shows that $T_{1,2}$ has properties analogous to those of the test statistic $T_{1,2}^*$ in Proposition \ref{prop1}, as if it were derived from the oracle residuals.

\begin{assumption}\label{assp_sigma}
Assume $Z \in \mathbb{R}^D$ is a sub-Gaussian vector with zero mean. That is, there exists a universal constant $\sigma >0$ such that  $E(e^{ t\mu^TZ})\leq e^{t^2\sigma^2/2}$ for any $t\in \mathbb{R}$ and $\mu\in \mathbb{R}^D$ with $\|\mu\|_2=1$.
\end{assumption}

\begin{assumption}\label{assp_sparsity} 
Consider representation of $\Omega$ using $m\times m$ blocks,
that is $\Omega=(\Omega_{\mathcal{G}_i,\mathcal{G}_j})$. Assume $\Omega$ is block-sparse, so that we have
$\max_i \sum^m_{j=1} \mathds{1}\{\Omega_{\mathcal{G}_i,\mathcal{G}_j}\neq 0\}\leq s$.
Also, assume that $s N^{-1/2}\log D=o(1)$.
\end{assumption}

\begin{assumption}\label{assp_eigen}
  The smallest and largest eigenvalues of $\Omega$, $\lambda_{\min}(\Omega)$ and $\lambda_{\max}(\Omega)$, are bounded that $0<1/c_*\leq \lambda_{\min}(\Omega)\leq \lambda_{\max}(\Omega)\leq c_*<\infty$ for universal constant $c_*>0$. 
\end{assumption}

\begin{assumption}\label{assp_groupsize}
For each $i'=1,\dots,m$, $d_{i'}:=|\mathcal{G}_{i'}|<d_*$ for universal constant $d_*>0$.
\end{assumption}

Assumptions \ref{assp_sigma} and \ref{assp_eigen} are commonly adopted in high-dimensional graphical model analysis (\cite{ren2015asymptotic, jankova2017honest}). The multivariate sub-Gaussian assumption is imposed to facilitate the theoretical analysis of our estimator but can be difficult to verify in practice. Nevertheless, our method is expected to remain reliable across a wide class of distributions, as demonstrated by our simulation studies, where categorical data are generated from truncated normal distributions, permutations of real data, and multinomial distributions (see Section \ref{sec:simulation} for more details). Unlike many existing studies, the precision matrix in our setting has a block structure, and we assume that the block size remains bounded in Assumption \ref{assp_groupsize} as guided by the MSA Data. Consequently, we impose the sparsity-related Assumption \ref{assp_sparsity} based on the block structure. The sample size requirement in Assumption \ref{assp_sparsity} aligns with conditions frequently used for high-dimensional inference (\cite{ren2015asymptotic, jankova2017honest}).

\begin{theorem}\label{thm1}
Under Assumptions \ref{assp_sigma}-\ref{assp_groupsize} and the null $H_0:\mathbf{P}^{(1,2)}=0$, we have that $$\lim_{N\to \infty} \displaystyle{\frac{\mathbb{P}(T_{1,2}\leq x)}{\mathbb{P}(\sum_{\ell'=1}^{d_1d_2} \wh{\Lambda}_{\ell'} W_{\ell'} ^2\leq x)}}=1,$$ where $W_{\ell'}\text{'s}\sim N(0,1)$ are independent, and $\wh{\Lambda}_{\ell'}\text{'s}$ are the eigenvalues of $\wh{\Cov(\mathbf{Q})}$ for $\ell'=1,\dots,d_1d_2$.
\end{theorem}

We reject the null hypothesis and classify the pair $(1,2)$ as a contact if $T_{1,2}>q^{(1,2)}_{0.05}$, where $\mathbb{P}(\sum_{\ell'=1}^{d_1d_2} \wh{\Lambda}_{\ell'} W_{\ell'} ^2> q^{(1,2)}_{0.05})=0.05$. The test statistic proposed in Theorem \ref{thm1} is a modified version of Wilk's lambda, where we replace the oracle errors with fitted residuals. For the CCA-related testing problem, several other spectrum-based test statistics exists, including the Hotelling-Lawley trace, Pillai trace, and Roy's largest root. See, for example, \cite{muller1984practical} for more details. These alternative statistics can also be used to test the same null hypothesis described in Proposition \ref{prop1} and are constructed using the sample canonical correlation coefficients. However, based on the simulations and preliminary results involving several protein families, we found that their performance closely resembles that of our modified Wilks' lambda statistic. Therefore, this article focuses primarily on the modified test statistic based on Wilk's lambda.

It is worth noting that given $m$ positions, there are $m(m-1)/2$ distinct test statistics $T_{i',j'}$'s, so the problem naturally falls within the framework of simultaneous hypothesis testing. This motivates the use of a multiple testing procedure with FDR control. As shown in Theorem 1, these test statistics may have different limiting distributions. To facilitate the construction of an FDR procedure, following \cite{xia2018multiple}, we apply a normalizing transformation and construct the procedure based on the transformed statistics. Under the assumption that the transformed statistics are not overly correlated, an assumption also adopted in \cite{liu2013gaussian} and \cite{xia2018multiple}, the FDR can be shown to be asymptotically controlled at the desired level. Due to the space limitation, a comprehensive study, including the full procedure, assumptions, and proofs, is provided in the supplementary material.

\begin{remark}
Consistent recovery of the underlying contact graph is also attainable under a minimal signal strength assumption. Let $\widehat{\mathcal{H}}_1=\{1\leq i'<j'\leq m: T_{i',j'}\geq K\log m\}$ denote the estimated set of contact pairs, where $K$ is a sufficiently large constant. Assuming that all true contact pairs satisfy a suitable minimal signal condition, arguments similar to those in \cite{ren2015asymptotic} imply that both false positives and false negatives vanish asymptotically. Specifically, $\mathbb{P}(\sum_{(i',j')\in\mathcal{H}_0}\mathds{1}\left\{T_{i',j'}\geq K\log m\right\}\geq 1)\to 0$ and $\mathbb{P}\left( \exists (i',j')\in\mathcal{H}_1: T_{i',j'}\leq K\log m \right)\to 0$, where $\mathcal{H}_0$ and $\mathcal{H}_1$ denote the sets of non-contact pairs and contact pairs, respectively. Thus, the contact graph can be consistently recovered with high probability.    
\end{remark}

\subsubsection{Comparison with other test statistics ($\ell_2$- and $\ell_{\infty}$-norm-based) and approaches (MI and PSICOV)} \label{sec:other stats}

There are other types of test statistics used for testing hypotheses related to covariance or precision matrices, which can be adapted for contact prediction and mutation analysis in our context with one-hot encoded data. For example, \cite{xia2018multiple} proposed a test statistic based on the $\ell_2$-norm of the self-normalized sample cross-covariance matrix to examine whether a specific submatrix of a Gaussian precision matrix is equals zero. Similarly, \cite{cai2013two} focused on the self-normalized sample covariance matrix but introduced a test statistic based on the $\ell_{\infty}\text{-norm}$.

Here, we present the constructions of $\ell_2$- and $\ell_{\infty}$-norm-based test statistics and their reference distributions. Given data $\mathbf{a}=(a_1,\dots,a_N)^T$ and  $\mathbf{b}=(b_1,\dots,b_N)^T$, let $\bar{a}$ and $\bar{b}$ denote the sample means of $\mathbf{a}$ and $\mathbf{b}$, respectively. Define the sample covariance between $\mathbf{a}$ and $\mathbf{b}$ as $\wh{\sigma}_{ab}=N^{-1} \sum_{k=1}^N (a_k-\bar{a})(b_k-\bar{b})$, and denote $\wh{\theta}_{ab}=N^{-2}\sum_{k=1}^N[(a_k-\bar{a})(b_k-\bar{b})-\wh{\sigma}_{ab}]^2$ the estimated variance for $\wh{\sigma}_{ab}$. The self-normalized sample covariance between $\mathbf{a}$ and $\mathbf{b}$ is then defined as $\wh{\sigma}_{ab}(\wh{\theta}_{ab})^{-1/2}$. For our testing problem, we focus on the pair $(i,j)=(1,2)$. The sample covariance matrix $\wh{\mathbf{S}}$ is computed from the residuals $(\wh{\mathbf{E}}_{(1,\backslash\{1,2\})},\wh{\mathbf{E}}_{(2,\backslash\{1,2\})})$ and is partitioned into four blocks, with $\wh{\mathbf{S}}_{12}$ denoting the $d_1\times d_2$ sample cross-covariance matrix and $\check{\mathbf{S}}_{12}$ representing its entry-wise self-normalized version. Subsequently, we modify the procedure from \cite{xia2018multiple} by defining the $\ell_2\text{-norm}$ based test statistic as $T_{1,2}^{\ell_2}=\|\check{\mathbf{S}}_{12}\|^2_2$, whose intermediate reference distribution follows a weighted Chi-squared distribution. The weights correspond to the eigenvalues of the covariance matrix of $\{(\check{\mathbf{S}}_{12})_{t_1,t_2},t_1=1,\dots,d_1;t_2=1,\dots,d_2\}$. Each entry of this matrix is computed analogously to $\wh{\theta}_{ab}$. Similarly, adapting the approach from \cite{cai2013two}, the $\ell_{\infty}\text{-norm}$ based test statistic is defined as $T_{1,2}^{\ell_{\infty}}=\|\check{\mathbf{S}}_{12}\|^2_{\infty}-4\log(d_2)+\log\log(d_2)$, where $\|\bM\|_{\infty} = \max_{t_1,t_2} |M_{t_1,t_2}|$ for matrix $\bM$. Its limiting distribution is Gumbel with a cumulative distribution function (CDF) given by $\exp(-\exp(-x/2)/(\sqrt{8\pi}))$.

In addition to the two test statistics mentioned above, we also include two non-inference-based approaches: PSICOV and MI. MI is among the most commonly used methods in contact prediction studies. See, for example, \cite{dunn2008mutual} for calculating MI between any two positions in MSA data. Although PSICOV is not a hypothesis test-based method, it uses precision matrices, which partially aligns with the principles of our approach. Detailed calculations of the scores used in PSICOV are provided in Section \ref{sec12}. Through both simulations and real data applications, we will demonstrate that our spectrum-based statistic consistently outperforms these four methods. This, in turn, validates the applicability of our approach to contact prediction and mutation analysis.

\subsection{Identifying amino acid combinations relevant to contacts and mutations}\label{sec:amino acid level}
The newly proposed approach not only allows inference about contact residues using fitted residuals but also enables identifying which combinations of amino acids are significantly partially correlated. For instance, consider the pair $(i,j)=(1,2)$ in a protein family that form a contact. One may further identify which combinations of amino acids yield strong correlation between these two positions. If we have access to oracle errors, $(\mathbf{E}_{(1,\backslash\{1,2\})},\mathbf{E}_{(2,\backslash\{1,2\})})$, such inferences become straightforward. Specifically, let $$\rho_{t_1,t_2}=\frac{\sum_{k=1}^N [\mathbf{E}_{(1,\backslash\{1,2\})}]_{k,t_1}[\mathbf{E}_{(2,\backslash\{1,2\})}]_{k,t_2}}{\sqrt{\sum_{k=1}^N [\mathbf{E}_{(1,\backslash\{1,2\})}]^2_{k,t_1}\sum_{k=1}^N [\mathbf{E}_{(2,\backslash\{1,2\})}]^2_{k,t_2}}}$$ denote the partial correlation based on oracle errors for any combination of amino acids $(t_1,t_2), t_1=1,\dots,d_1; t_2=1,\dots,d_2$, within the pair of positions $(i,j)=(1,2)$. It is well-known that $\rho_{t_1,t_2}$ is asymptotically normal. Let $\wh{\rho}_{t_1,t_2}$ be the counterpart computed using the residuals. As shown in Proposition \ref{prop02}, under the conditions $s\log D/\sqrt{N}=o(1)$ and other regularity assumptions, $\wh{\rho}_{t_1,t_2}$ and $\rho_{t_1,t_2}$ are asymptotically equivalent.

\begin{proposition}\label{prop02}
Under Assumptions \ref{assp_sigma}-\ref{assp_groupsize}, we have that $\sqrt{N}(\rho_{t_1,t_2}-\wh{\rho}_{t_1,t_2})=o_p(1)$. Thus, if $[\mathbf{P}^{(1,2)}]_{t_1,t_2}=0$, then $\sqrt{N}\wh{\rho}_{t_1,t_2}\overset{\mathcal{L}}{\to} N(0,\gamma^2)$, where $$\gamma^2=\frac{\Var([\mathbf{E}_{(1,\backslash\{1,2\})}]_{\ell,t_1}[\mathbf{E}_{(2,\backslash\{1,2\})}]_{\ell,t_2})}{\Var([\mathbf{E}_{(1,\backslash\{1,2\})}]_{\ell,t_1})\Var([\mathbf{E}_{(2,\backslash\{1,2\})}]_{\ell,t_2})}.$$ 
\end{proposition}

However, $\gamma^2$ in Proposition \ref{prop02} is unknown. Therefore, in practice we have to replace it by a consistent estimator, and use $$\sqrt{N}\wh{\rho}_{t_1,t_2}/\wh{\gamma}=\frac{\sum_{k=1}^N [\wh{\mathbf{E}}_{(1,\backslash\{1,2\})}]_{k,t_1}[\wh{\mathbf{E}}_{(2,\backslash\{1,2\})}]_{k,t_2}}{\sqrt{\sum_{k=1}^N [\wh{\mathbf{E}}_{(1,\backslash\{1,2\})}]^2_{k,t_1}[\wh{\mathbf{E}}_{(2,\backslash\{1,2\})}]^2_{k,t_2}}}$$ as the test statistic. The reference limiting distribution will still be the standard normal, and we may reject the null hypothesis if $|\sqrt{N}\wh{\rho}_{t_1,t_2}/\wh{\gamma}|>z_{0.025}$, where $z_{0.025}$ is the $0.025$ upper-quantile of the standard normal distribution. Note that when considering a pair of two positions in contact, there can be up to $20\times 20=400$ tests of partial correlations at the amino acids level. A practical approach is to set a cutoff value for $p$-values and select the top-ranked amino acid combinations with $p$-values below this threshold.

\section{Implementation} 
\label{sec3} 

To summarize our test, we outline its implementation for a given protein family with $N$ sequences and $m$ positions in Algorithm \ref{althm1}. For convenience of reference, we refer to our method as CATParc (\textbf{CAT}egorical \textbf{Par}tial \textbf{C}orrelation) throughout the remainder of the paper. For each pair of positions, to acquire the two residual matrices, it demands running the multivariate group Lasso for $m(m-1)$ times. Thus, for large protein families with many positions, the computational cost can be notably high. To mitigate this, after performing the one-versus-rest regressions in Step 1 of Algorithm \ref{althm1}, for any pair $(i,j)$ satisfying certain conditions outlined in Step 2, we directly use the fitted residuals from the one-versus-rest regressions as  $(\wh{\mathbf{E}}_{(i,\backslash\{i,j\})},\wh{\mathbf{E}}_{(j,\backslash\{i,j\})})$  without further computation. As noted in \cite{ren2015asymptotic}, this strategy yields exact estimates of the regression residual matrices for each pair $(i,j)$, rather than an approximation. Also, it reduces the number of multivariate group Lasso from $m(m-1)$ to the order of $sm$. More discussion can be found in the supplementary material on the discussion of one-versus-rest regressions.

\begin{algorithm}[!h]
\caption{~Implementation of CATParc}
\label{althm1}
{\small
\begin{enumerate}[label=\textbf{Step \arabic*:},itemindent=1cm]
\item One-versus-rest regressions
  \begin{enumerate}[label=(\roman*)]
 \item   For each $i\in \{1,2,\dots,m\}$, regress $\mathbf{X}_{\boldsymbol{\cdot},\mathcal{G}_i}$ against $\mathbf{X}_{\boldsymbol{\cdot},\mathcal{G}^c_i}$ using multivariate group Lasso as below, where $\mathcal{P}_{i}^c= \{1,2,\dots,m\}\setminus\{i\}$,
 $$\wh{\mathbf{B}}_{(i,\backslash i)}=\argmin\nolimits_{\breve{\mathbf{B}}\in \mathbb{R}^{(p-d_i)\times d_i}} (2N)^{-1}\|\mathbf{X}_{\boldsymbol{\cdot},\mathcal{G}_i}-\mathbf{X}_{\boldsymbol{\cdot},\mathcal{G}^c_{i}}\breve{\mathbf{B}}\|_2^2+\sum_{i'\in \mathcal{P}_{i}^c} \lambda_{i'}\|\breve{\mathbf{B}}_{g_{i'}}\|_2.$$
 \vspace{-0.25cm}
 \item Obtain the residual matrix $\wh{\mathbf{E}}_{(i,\backslash i)}$. 
   \end{enumerate}
\item  For each pair $(i,j)$,
 \begin{enumerate}[label=\textbf{2.\arabic*}, itemindent=0cm]
 \item{\bf Compute pairwise statistics}
 \begin{enumerate}[label=(\roman*)]  \item If $\|(\wh{\mathbf{B}}_{(i,\backslash i)})_{g_j}\|_2=0$,  let $\wh{\mathbf{E}}_{(i,\backslash\{i,j\})}=\wh{\mathbf{E}}_{(i,\backslash i)}$. Otherwise, solve \eqref{objective} to get $\wh{\mathbf{E}}_{(i,\backslash\{i,j\})}$. Obtain  $\wh{\mathbf{E}}_{(j,\backslash\{i,j\})}$ similarly.
 \item Without loss of generality, assume $d_i\leq d_j$. Compute $\wh{r}_1,\dots,\wh{r}_{d_i}$ based on $(\wh{\mathbf{E}}_{(i,\backslash\{i,j\})},\wh{\mathbf{E}}_{(j,\backslash\{i,j\})})$ and  obtain  $T_{i,j}=-N\log\prod_{\ell=1}^{d_i}(1-\wh{r}^2_{\ell})$.
 \end{enumerate}
\item{\bf Inference-based detection}
 \begin{enumerate}[label=(\roman*)]
 \item Compute the $p$-value by referencing the associated weighted Chi-squared distribution\footnotemark and conclude that $(i,j)$ form a contact if $p$-value exceeds $0.05$. 
  \item For the pairs of interest, compute the normalized partial correlation in Section \ref{sec:amino acid level} for each amino acid combinations, and then get the associated $p$-value by referencing to $N(0,1)$.
\end{enumerate}
\end{enumerate} 
\end{enumerate}
}
\end{algorithm}

As shown in Theorem D.1 of the supplementary materials, the regularization parameter $\lambda_{i'}$ is set to be $C(\sqrt{d_{i'}d_i/N}+\sqrt{A\log(m-2)/N})$, with universal constants $A > 2$ and $C > 0$, similar to \cite{mitra2016benefit}. In practice, cross-validation can be used to determine appropriate values for $(A, C)$. However, this process is time-consuming as the parameters must be tuned separately for each pair $(i,j)$. To address this, we follow \cite{mitra2016benefit} by fixing $A = 2$. As for $C$, we tune it on a randomly selected subset of one-versus-rest regressions and then use the median of the tuned values for all pairs. Preliminary trials on different protein families show that choosing $C$ within the range $(0.05, 0.1)$ does not significantly affect the results. A sensitivity analysis examining how the choice of $C$ affects the performance of our method is conducted on two representative protein families. The results support this observation and are provided in the supplementary material. Thus, unless specified otherwise, we adopt $(A, C) = (2, 0.07)$ for subsequent simulations and real data analysis.

\footnotetext{The package \href{https://github.com/FanY4098/CATParc}{\texttt{CATParc}} provides $p$-values based on the weighted Chi-squared distribution.}

\section{Numerical Experiments}\label{sec:simulation}

For the simulation studies, we consider three generating mechanisms for multivariate categorical data: (1) truncating continuous random variables (Cases 1-4, \cite{jernigan2021large}), 
(2) synthesizing data from MSA data to better mimic real-world conditions (Cases 5-8), and (3) sampling from multinomial distributions (Cases 9-10). We compare our method, CATParc, with two inference-based procedures that use either $\ell_2$- or $\ell_{\infty}$-norm statistics, as detailed in Section \ref{sec:other stats}. For a comprehensive comparison, we also include PSICOV, although it was not originally designed for inference. Receiver operating characteristic (ROC) curves are used to assess the performance. Due to space limitations, the detailed simulation setups and the results for Cases 1-4 and Cases 9-10 are included in the supplementary material. In addition, we present simulation results with histograms to verify Theorem~\ref{thm1}, with further details also provided in the supplementary material.

As described in Section \ref{sec1.1}, let $m$ be the total number of positions along the protein sequence. In each simulated case, these positions are divided into $u$ groups of equal size, with $h\coloneqq m/u$ positions per group. The data are generated such that positions within the same group are partially correlated, while positions across different groups are independent and, thus, partially uncorrelated.

\subsection{Simulations based on the real data}

Permuting real data is a straightforward way to simulate protein sequences that mimic the real biological conditions. Specifically, for an MSA data, we first divide the adjacent positions into several groups. When the group size, $h$, is small, pairs of positions within the same group can be roughly considered as contacts, given their relatively close spatial locations. We then randomly permute the rows of different groups, so that the dependence across groups is eliminated while the dependence within groups is preserved. As a result, any two positions from different groups are considered non-contact residues.

\begin{figure}[h!]
\centering
\begin{subfigure}{.4\linewidth}
  \includegraphics[width=\linewidth]{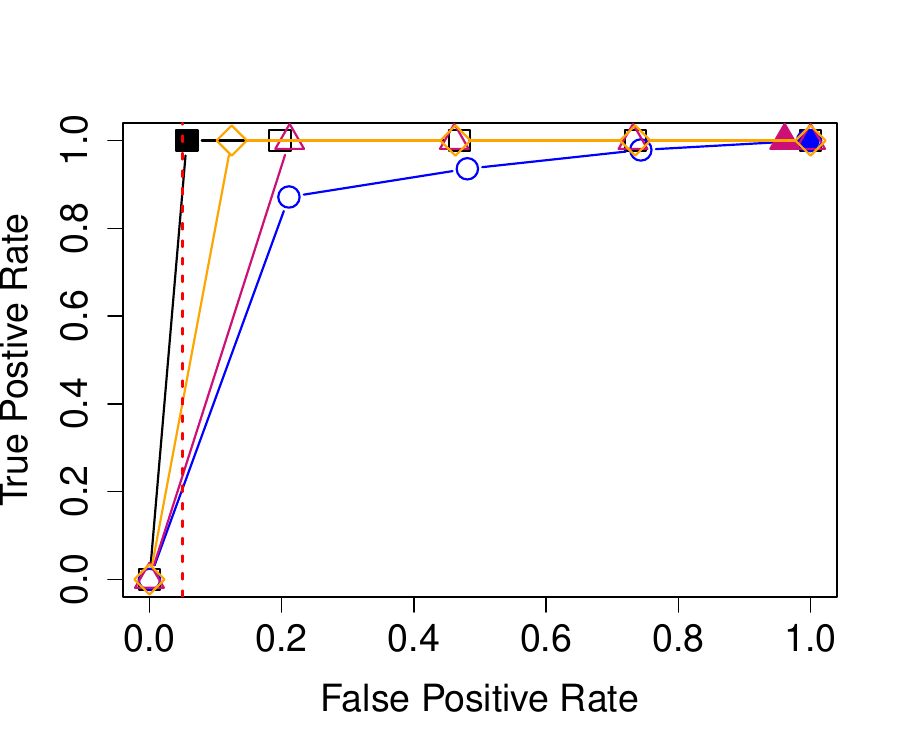}  
  \caption{Case 5}
\end{subfigure}
\begin{subfigure}{.4\linewidth}
  \includegraphics[width=\linewidth]{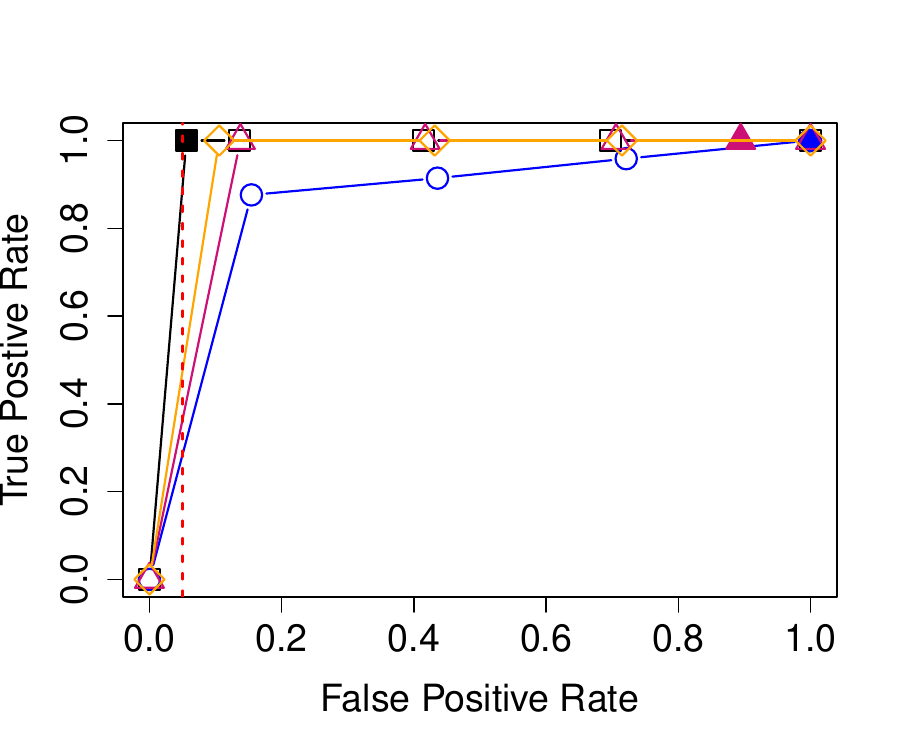}  
  \caption{Case 6}
\end{subfigure}

\begin{subfigure}{.4\linewidth}
  \includegraphics[width=\linewidth]{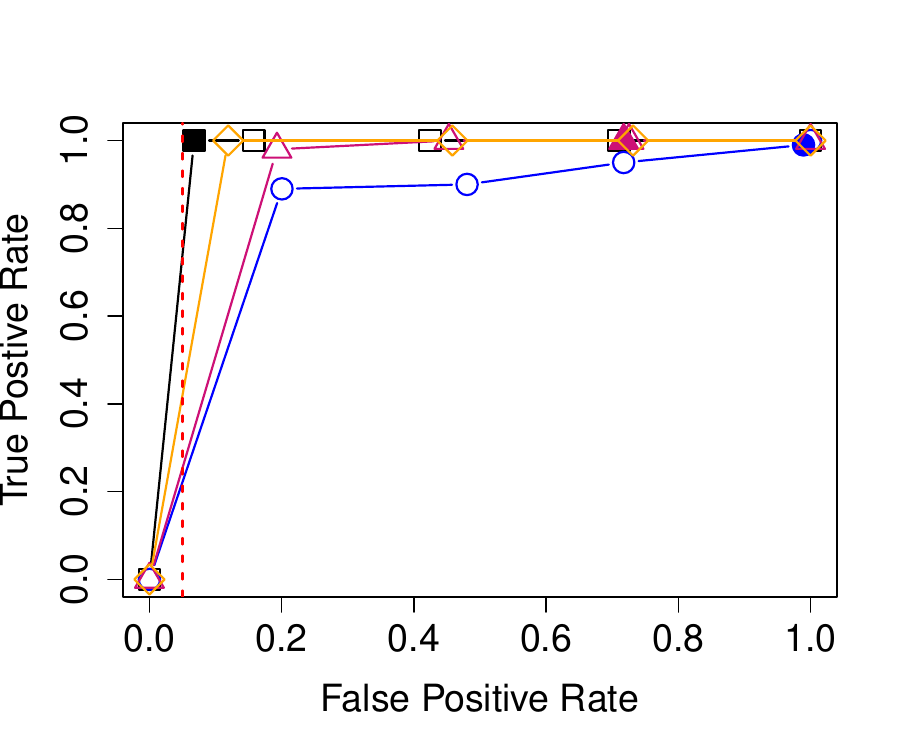}  
  \caption{Case 7}
\end{subfigure}
\begin{subfigure}{.4\linewidth}
  \includegraphics[width=\linewidth]{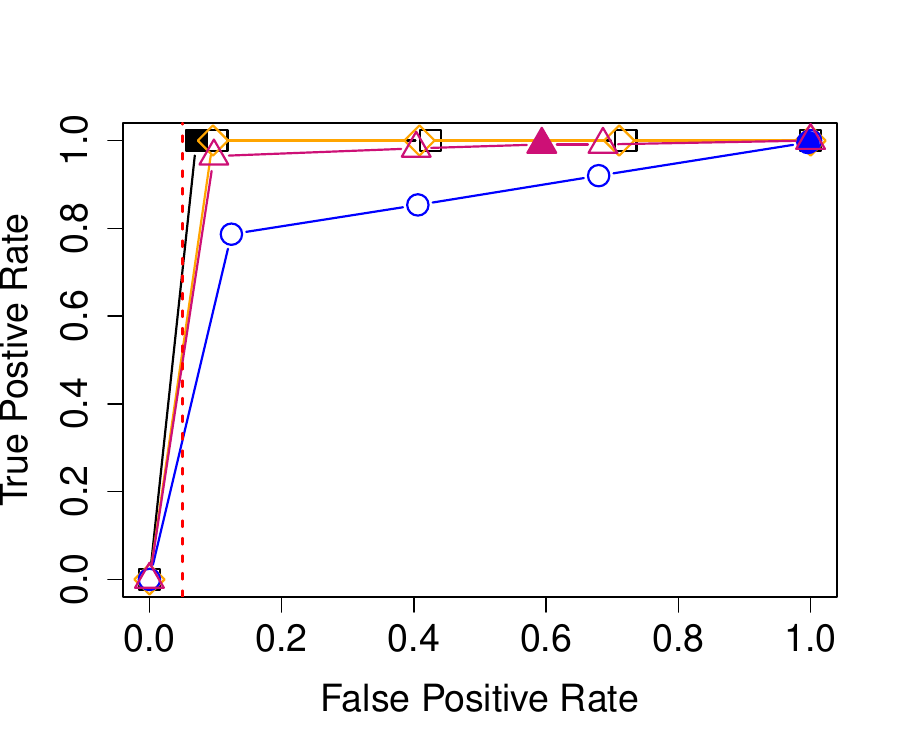}  
  \caption{Case 8}
  \end{subfigure}
\caption{ROC curves for Cases 5-8: $\square$, $\bigcirc$, $\bigtriangleup$, and $\lozenge$ represent our method (CATParc), L2, SUP, and PSICOV respectively. The red dashed line indicates a Type I error of $0.05$, and the solid colored symbols mark the empirical Type I error and power when the nominal level is $0.05$.}
\label{fig:roc2}
\end{figure}

In MSA data, some amino acids appear in very small or very large proportions at certain positions, which results in skewed binary columns in the one-hot encoded data. These skewed columns can impair the performance of subsequent tests. To address this, we introduce a trimming process: for each position in the MSA data, we compute the proportions of all amino acids. If any amino acid has a proportion below a certain threshold, the entire sequence containing that amino acid at that position is removed.

The simulated data for Cases 5-8 below are also based on PF00188, which originally has $25,686$ sequences and $118$ positions. For each case, we retain specific positions, divide them into $u$ groups, and randomly permute the rows within each group. In Case 5, positions $25$-$94$ are retained, and the above trimming process is applied with a threshold value $0.005$, which results in a data set with $14,404$ sequences. Then, every $5$ positions form a group, giving $u=14,h=5$. In Case 6, the same positions and trimming threshold as Case 5 are used, but groups are formed with every $10$ positions so that $u=7,h=10$. For Case 7, positions $35$-$84$ are retained and a trimming threshold of $0.01$ is applied, which yields a data with $7,911$ sequences. Groups are formed with every $5$ positions so that $u=10,h=5$. Case 8 follows the same setup as Case 7 except groups are formed with every $10$ positions so that $u=5,h=10$. All experiments are repeated for $50$ times.

To display the ROC curves in Figure \ref{fig:roc2}, we take the median false positive rates and true positive rates across the $50$ replicates for each fixed cut-off. Vertical dashed lines indicate a Type I error of $0.05$, while the solid symbols, whenever applicable, mark the Type I error and empirical power for different test statistics when the nominal Type I error is $0.05$. As shown in Figure \ref{fig:roc2}, CATParc not only maintain a marginal advantage over PSICOV across Cases 5-8 but, more importantly, produces valid $p$-values to quantify uncertainty, which makes it more informative. In Cases 7 and 8, the Type I error rate for CATParc is slightly inflated, which may be due to misspecifications of the null and alternative. Specifically, while we consider all pairs within the same group as partially correlated, some may actually be distant in the 3-dimensional structure and should be instead treated as partially uncorrelated.

\section{Predicting protein residue contacts}\label{sec:real}
\subsection{Contact prediction}\label{sec::contact prediction}

For a comprehensive assessment of our proposed CATParc, in Section  \ref{sec:main thm} for contact prediction across diverse protein fold types, we collect six protein families representing three structural categories: alpha-helical, beta-sheet, and mixed. Within each category, we include one large and one small protein family to account for different size ranges. Detailed descriptions of these six families are provided in the supplementary materials. The MSA data are collected from the Pfam Database, and residue contact information is derived from the experimentally determined structures of each protein family. Residues are considered in contact if the spatial distance between their $C\text{-}\beta$ atoms is within $4.5$ $\mathring{A}$.

\begin{table}[!ht]
\centering
\caption{AUC values of different methods for contact prediction}
\scalebox{0.8}{
\begin{tabular}{|m{3.5cm}|cc|cc|cc|}
\hline
Protein types  & \multicolumn{2}{c|}{Mixed alpha-beta}  & \multicolumn{2}{c|}{Alpha-helical}     & \multicolumn{2}{c|}{Beta-sheet}        \\ \hline\\[-1em]
\diagbox[width=3.6cm, trim=r]{\theadfont Methods}{\theadfont Protein family} & \multicolumn{1}{c|}{PF01037} & PF01430 & \multicolumn{1}{c|}{PF13560} & PF00502 & \multicolumn{1}{c|}{PF02115} & PF02823 \\ \hline \\[-1em]
CATParc          & \multicolumn{1}{c|}{0.827}   & 0.834   & \multicolumn{1}{c|}{0.912}   & 0.859   & \multicolumn{1}{c|}{ 0.800}   & 0.799   \\ \hline \\[-1em]
SUP           & \multicolumn{1}{c|}{0.693}   & 0.683   & \multicolumn{1}{c|}{0.756}   & 0.597   & \multicolumn{1}{c|}{ 0.621}   & 0.683   \\ \hline \\[-1em]
L2           & \multicolumn{1}{c|}{0.650}   & 0.634   & \multicolumn{1}{c|}{0.780}   & 0.597   & \multicolumn{1}{c|}{ 0.589}   & 0.642   \\ \hline \\[-1em]
MI          & \multicolumn{1}{c|}{0.656}   & 0.657   & \multicolumn{1}{c|}{0.743}   & 0.676   & \multicolumn{1}{c|}{ 0.678}   & 0.630   \\ \hline \\[-1em]
PSICOV         & \multicolumn{1}{c|}{0.778}   & 0.795   & \multicolumn{1}{c|}{0.845}   & 0.667   & \multicolumn{1}{c|}{ 0.694}   & 0.784   \\ \hline \\[-1em]
GREMLIN            & \multicolumn{1}{c|}{0.705}   & 0.734   & \multicolumn{1}{c|}{0.726}   & 0.668   & \multicolumn{1}{c|}{0.620}   & 0.683   \\ \hline \\[-1em]
plmDCA             & \multicolumn{1}{c|}{0.748}   & 0.779   & \multicolumn{1}{c|}{0.800}   & 0.787   & \multicolumn{1}{c|}{0.705}   & 0.758   \\ \hline \\[-1em]
mfDCA             & \multicolumn{1}{c|}{0.659}   & 0.717   & \multicolumn{1}{c|}{0.764}   & 0.727   & \multicolumn{1}{c|}{0.741}   & 0.684   \\ \hline \\[-1em]
GaussDCA         & \multicolumn{1}{c|}{0.688}   & 0.781   & \multicolumn{1}{c|}{0.759}   & 0.729   & \multicolumn{1}{c|}{0.675}   & 0.723   \\ \hline \\[-1em]
ESM2         & \multicolumn{1}{c|}{0.633}   & 0.702   & \multicolumn{1}{c|}{0.647}   & 0.678   & \multicolumn{1}{c|}{0.721}   & 0.658   \\ \hline 
\end{tabular}
 }
\label{table_no apc}
\end{table}

For each family, we examine every pair of positions by computing the proposed test statistics and the corresponding $p$-values. We have found that using the Chi-squared distribution yields satisfactory results while reducing computational costs. Therefore, unless specified otherwise, we use the Chi-squared distribution as the limiting distribution in subsequent real data applications. For comparison, we also include Mutual Information (MI), PSICOV, test statistics based on the $\ell_2$- and $\ell_{\infty}$-norm, four variants of DCA, i.e., mfDCA (\cite{weigt2009identification}), plmDCA (\cite{ekeberg2013improved}), GREMLIN (\cite{kamisetty2013assessing}), and GaussDCA (\cite{baldassi2014fast}), and ESM2 (\cite{lin2023evolutionary}, as a representative of large language model–based methods) for a more comprehensive comparison with our proposed CATParc. Further details of $\ell_2$- and $\ell_{\infty}$-norm based test statistics can be found in Section \ref{sec:other stats}. It is worth noting that mfDCA, plmDCA, and GREMLIN are all based on the well-known Potts model but adopt different strategies for parameter estimation. mfDCA, the earliest implementation of DCA, approximates the inverse Potts model parameters by applying mean-field theory to the correlation matrix derived from the MSA data, whereas plmDCA and GREMLIN rely on pseudolikelihood maximization. GaussDCA, in contrast, replaces discrete amino acid variables with continuous Gaussian random variables. This approximation makes it computationally attractive. ESM2 does not require training a separate model for each protein family of interest. By simply providing a single protein sequence, ESM2 outputs a predicted contact matrix, where each entry denotes the estimated probability of two positions being in contact. However, ESM2 cannot exploit information from an MSA, as it only accepts a single sequence as input. We use the Area Under the Curve (AUC) to evaluate the accuracy of each method. For CATParc and the $\ell_2$-norm-based test, we compute the AUC based on the standardized test statistics. In contrast, for the $\ell_{\infty}$-norm-based test, we use the original test statistics, as they are already normalized and share the same limiting distribution. For non-inference-based methods like MI, PSICOV, DCA variants and ESM2, scores are generated for each pair of positions to quantify the strength of the direct coupling, and these scores are used to compute the AUC. In addition, to obtain the ESM2 results, for each protein family we first filtered out sequences with a gap proportion exceeding 1\%. The remaining sequences were then passed into ESM2 to generate contact predictions, and the AUC was computed based on each sequence; we report the best AUC obtained.

As shown in Table \ref{table_no apc}, CATParc consistently outperforms the others in terms of AUC across all protein families, demonstrating its broad applicability to different types of protein structures. 
Among all three categories of real data, CATParc performs best on alpha helix-only proteins and performs worst on beta sheet-only proteins. In general, the quality of the MSA plays a critical role in the accuracy of these predictions. However, structural features of the protein also influence prediction outcomes \citep{caporaso2008detecting}. Alpha-helical regions generally result in the most accurate contact predictions due to their regular and predictable structure. In an alpha helix, residues that are close in sequence (i.e., $i$ to $i+3$ or $i+4$) are typically in contact (short-range contacts) due to the helical turn. The overall high density of residue contacts may lead to strong coevolutionary signals, especially when they are well-conserved across homologs. In contrast, beta sheets are more challenging to predict because the length of the beta strands is highly variable and contacts often involve residues that are distant in sequence, which form long-range contacts. These contacts are intrinsically more difficult to detect. Mixed alpha-beta proteins offer an intermediate case. Their overall performance in our predictions tends to reflect a balance between the alpha helical contacts and the beta sheet contacts.

We have also carried out a benchmarking study using PF01037 and PF02115, where we draw subsamples of different sizes and evaluate both the prediction accuracy and computational efficiency of all methods across varying sample sizes. Due to space limitations, the related details and results are provided in the supplementary material.


\subsection{Identification of amino acids combinations contributing to the residual contacts}\label{sec::amcidcomp}

As described in Section \ref{sec:amino acid level}, a key aspect of our procedure, which departs from existing algorithms like PSICOV, is the ability to identify specific amino acid combinations within each detected contact pair that significantly contribute to the contact. This allows for a fine characterization of compensatory mutations. To demonstrate, we apply our procedure to the class A beta-lactamase family, which consists of $8,403$ protein sequences and $263$ positions in the MSA. This protein family is widely used as a benchmark for evaluating the effects of protein mutations \citep{stiffler2015evolvability, figliuzzi2016coevolutionary}. Its activity can be assessed by measuring organism survival in the presence of antibiotics, and previous studies have successfully analyzed residue-level coevolution in this family \citep{stiffler2015evolvability,meier2021language}.

\begin{figure}[h]
\centering
\scalebox{0.85}{
\begin{minipage}{.5\textwidth}
\centering
  \begin{subfigure}[b]{0.95\linewidth}
  \vspace{0.5cm}
  \includegraphics[width=\linewidth,height=9.5cm]{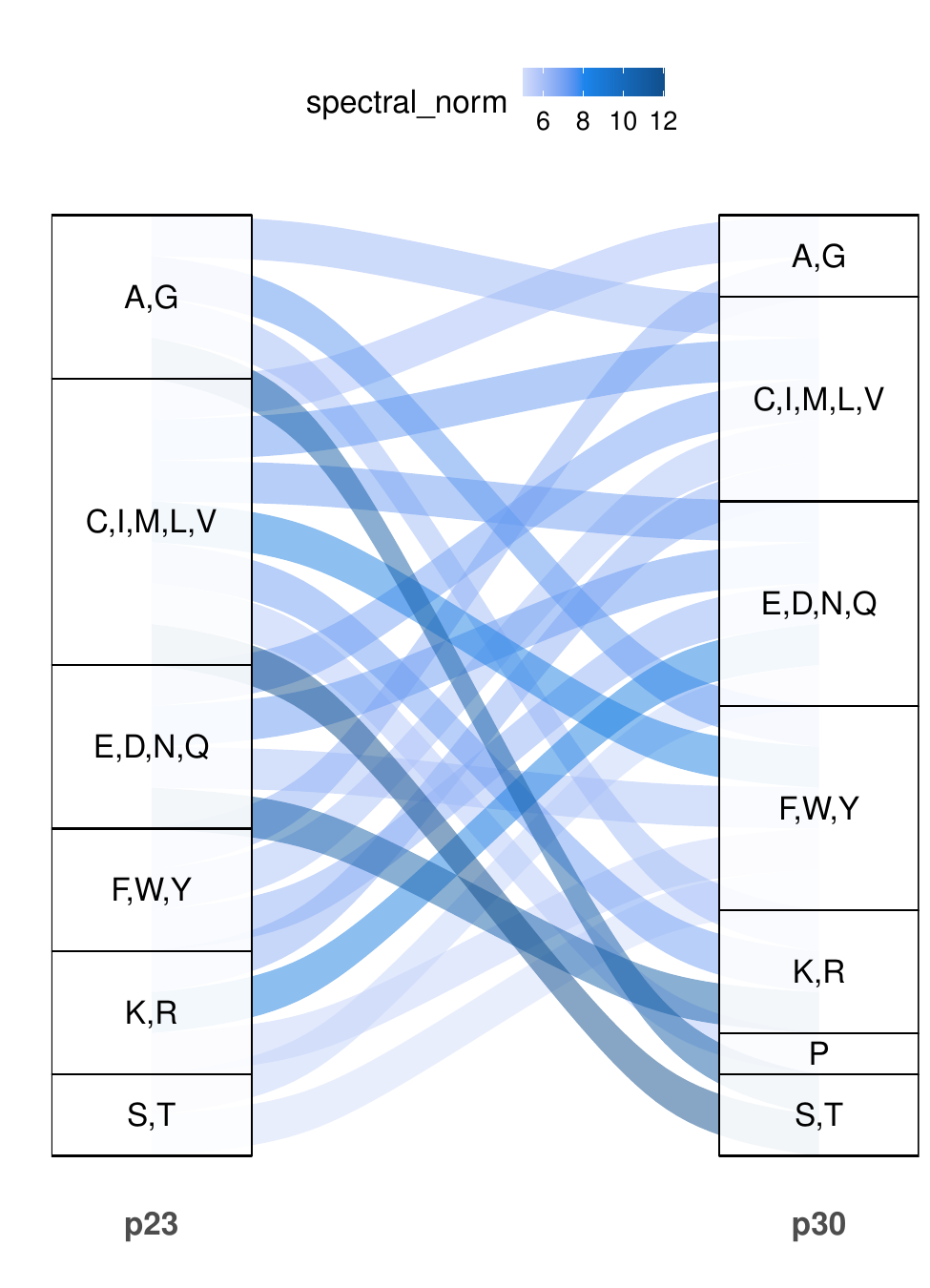}
  \caption{}
  \end{subfigure} 
\end{minipage}%
\hspace{.3 cm}
\begin{minipage}{.45\textwidth}
  \centering    
  \begin{subfigure}[b]{0.9\linewidth}
    \centering
    \includegraphics[width=0.9\linewidth, height=4.15cm]{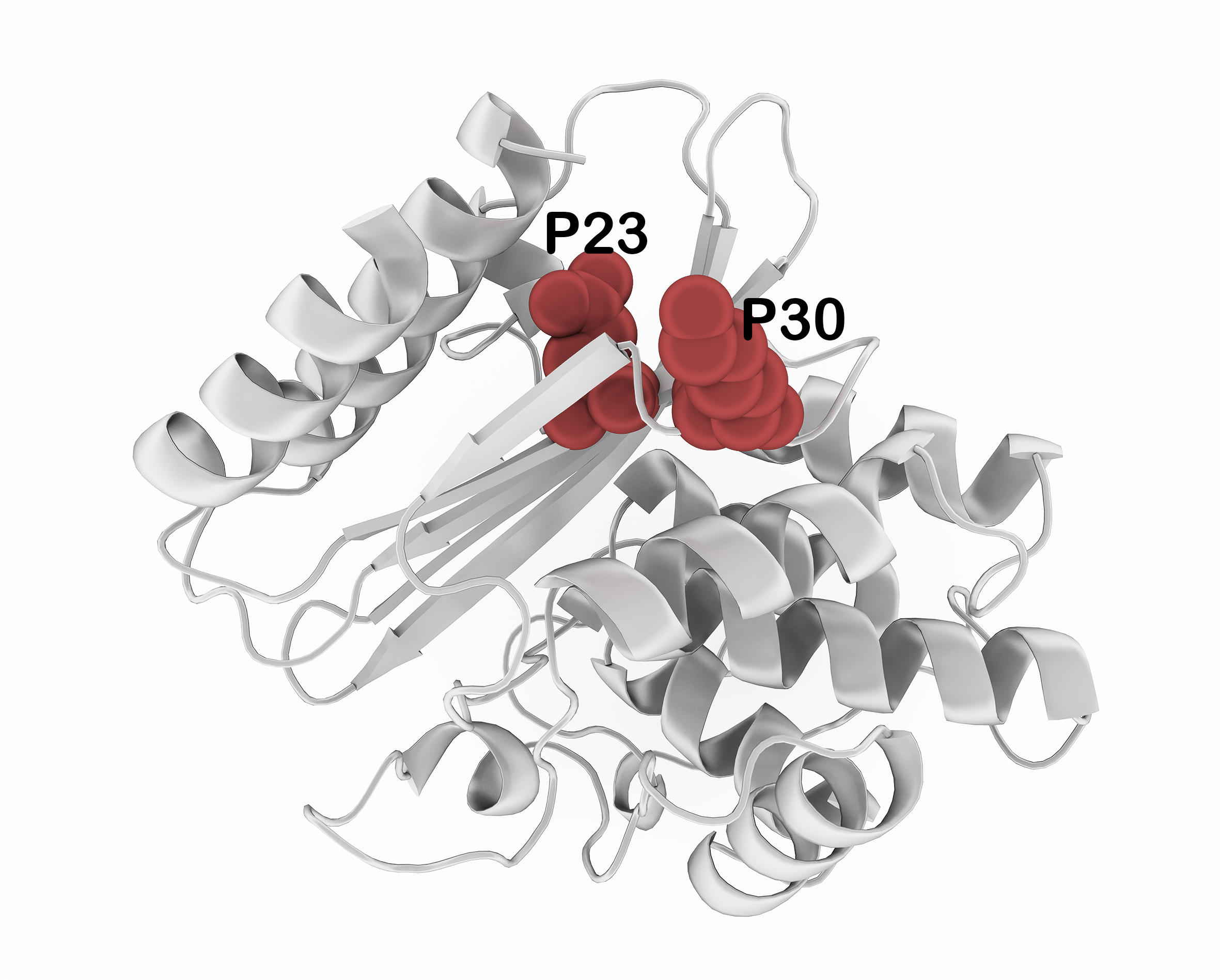}
    \caption{}
  \end{subfigure}\\ 
  \begin{subfigure}[b]{0.9\linewidth}
    \centering
    \includegraphics[width=\linewidth,height=5.15cm]{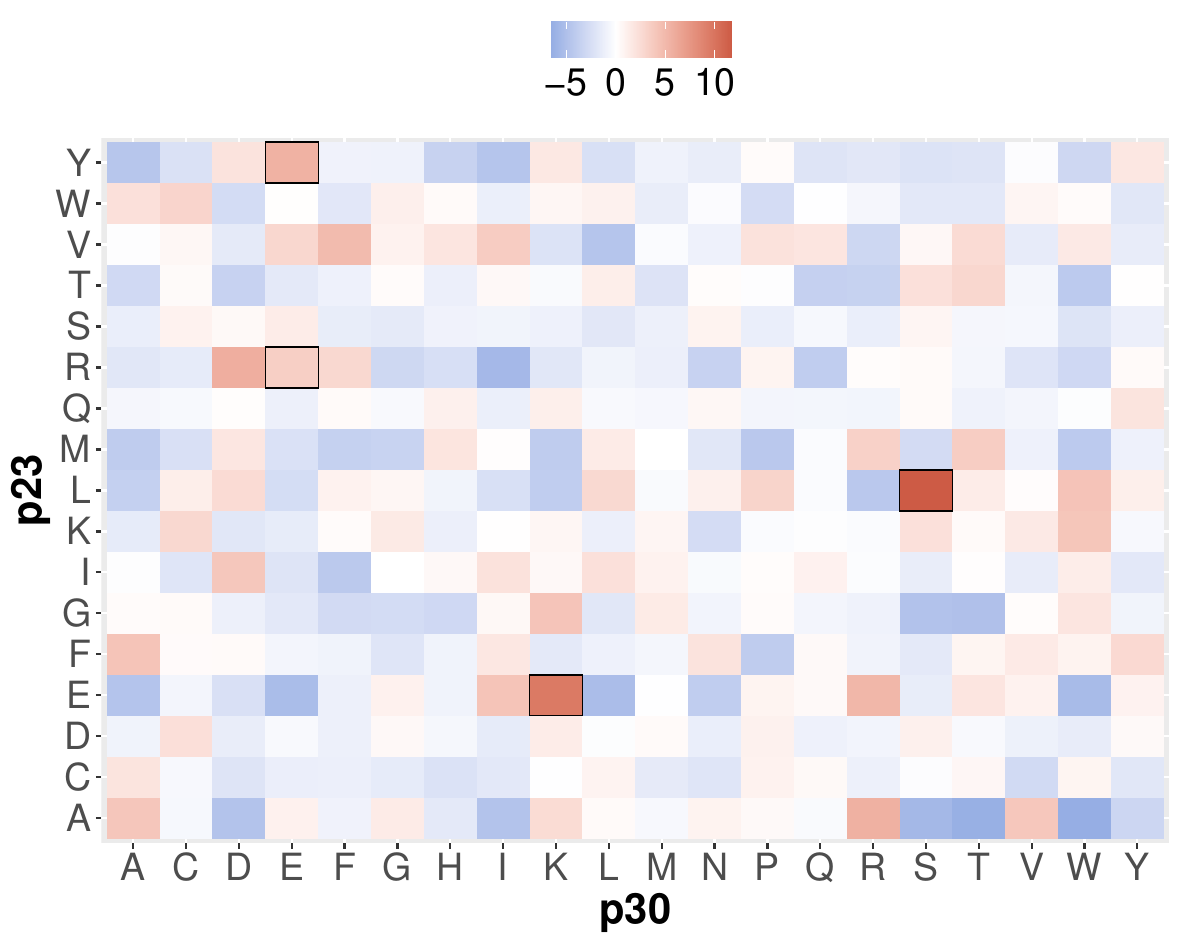}
    \caption{}
  \end{subfigure} 
\end{minipage}
}
\caption{(a) Sankey diagram showcasing the correlation of amino acid groups between positions $23$ and $30$. The strength of the correlation is measured by the spectral norm of the associated submatrix of the self-normalized partial correlation matrix. (b) Residues $(23,30)$ highlighted in the crystallization structure of beta-lactamase TEM. (c) Self-normalized partial correlation map of amino acid pairs between positions $23$ and $30$. } 
\label{fig:heatmap}
\end{figure}

We focus on positions $23$ and $30$, which are known to strongly correlate and form close contacts in the structure. As shown in Panel (b) of Figure \ref{fig:heatmap}, these two positions are indeed proximal in the protein’s crystallization structure, based on spatial distance. For this pair, there are $17\times 20=340$ amino acid combinations, for which we compute the normalized partial correlations. Following \citet{murphy2000simplified}, we group the $20$ amino acids into $8$ categories. For each pair of groups between the two positions, we calculate the spectral norm of the corresponding partial correlation submatrix to assess the strength of the coupling between them. In the Sankey diagram in Panel (b) of Figure \ref{fig:heatmap}, for clarity, only the top $23$ links are displayed. This visualization provides a detailed characterization of which groups of amino acids exhibit stronger correlations.

In Panel (c) of Figure \ref{fig:heatmap}, we present a partial correlation map that allows for a more detailed examination of the correlations between all amino acid combinations. The $y$-axis represents position $23$, and the $x$-axis represents position $30$, with the letters indicating the amino acids observed at each position. Only a few amino acid combinations show significant positive partial correlations, such as E(p23)-K(p30), Y(p23)-E(p30) (hydrogen bond), and R(p23)-E(p30) (charge swap). These findings are consistent with the sequence composition observed at these two positions in the multiple sequence alignment, which suggests that compensatory interactions between charged or polar residues may be conserved at positions $23$ and $30$. Other correlated pairs, like L(p23)-S(p30), do not exhibit clear compensatory interactions and are likely due to other forms of interaction such as higher-order mutation correlations, or a false alignment.

\section{Prediction of mutation effects}\label{sec::MutPred}

Mutation effect assessment involves evaluating how amino acid mutations affect the function of a protein \citep{liu2012sequence}. Single-point or multiple-point mutations can be experimentally introduced into a wild-type sequence to generate different mutants. The effect of these mutations on specific protein features, such as melting temperature or catalytic activity, can be assessed experimentally. Although experimental approaches provide valuable insights into the consequences of mutations, they are often costly and time-consuming.


Recent studies have shown promising results in predicting mutation effects using computational approaches. For example, unsupervised methods assign scores to mutant sequences based on the probabilistic model of the entire sequence and the corresponding wild-type sequence \citep{hopf2017mutation,figliuzzi2016coevolutionary}. These scores serve as proxies for experimental mutation effects. In contrast, supervised approaches, such as Evolutionary Scale Modeling (ESM), have been proposed when the experimental mutation effects are available \citep{rives2021biological}. ESM utilizes pre-trained language models to extract $1,280$ features from each mutant sequence, which are used to train algorithms such as $k$-nearest neighbors ($k$NN), support vector machines (SVM), or random forests (RF). Here, the experimental mutation effects and the generated features serve as responses and predictors, respectively. 

\subsection{New feature generation} Using the estimated partial covariance matrices from Section \ref{sec:test}, we demonstrate that our procedure generates complementary features that enhance the performance of ESM. Specifically, adding just two features from CATParc to the existing $1,280$ ESM features improves mutation effect prediction by $3\%$. To validate the effectiveness of our features, we also generate additional features using PSICOV for comparison.

We draw upon the idea in \cite{hopf2017mutation} to generate new features as follows. Let $\mathbf{\Sigma}^{(i,j)}=\Cov(\boldsymbol{\mathcal{E}}_{(i,\backslash(i,j))},\boldsymbol{\mathcal{E}}_{(j,\backslash(i,j))})$ denote the population partial covariance matrix between positions $(i,j)$, and define $\mathbf{C}_{ij}\coloneqq \wh{\mathbf{\Sigma}}^{(i,j)}$ as its empirical counterpart, derived from the residuals from procedures in Section \ref{sec: estimation procedure}. For a mutant sequence $\mathbf{a}=(a_1,\dots,a_m)$, we define $\mathbf{C}(\mathbf{a})=\sum_{i<j}\mathbf{C}_{ij}(a_i,a_j)$ to quantify the overall pairwise interactions within the given sequence, as $\mathbf{C}_{ij}(\cdot,\cdot)$ is the estimated partial covariance of two amino acids at positions $(i,j)$. Furthermore, for each position $i \in \{1,\dots,m\}$, define $\mathbf{M}_i(\cdot)=\sum_{j\neq i}\sum_{b\in \mathcal{A}}\mathbf{C}_{ij}(\cdot,b)$ where $\mathcal{A}$ is the set of twenty different amino acids. Here, $\mathbf{M}(\mathbf{a})=\sum_{i=1}^m\mathbf{M}_i(a_i)$ measures the marginal favorability of $\mathbf{a}$ across $m$ positions. Let $\mathbf{w}=(w_1,\dots,w_m)$ represent the wild-type sequence, for any mutant sequence $\mathbf{a}$, we define the {\it two new features generated using our approach} as $\Delta\mathbf{C}(\mathbf{a})=\mathbf{C}(\mathbf{a})-\mathbf{C}(\mathbf{w})$ and $\Delta\mathbf{M}(\mathbf{a})=\mathbf{M}(\mathbf{a})-\mathbf{M}(\mathbf{w})$. Similarly, we can generate two additional features based on PSICOV by replacing $\mathbf{C}_{ij}$ with the submatrix of the estimated precision matrix corresponding to the pair $(i,j)$.

\subsection{Numerical evidences} As a demonstration, we analyze mutations in the Beta-lactamase enzyme family and use the experimentally determined mutation effect data from \cite{hopf2017mutation} as a testing benchmark. A total of $5,397$ mutated sequences are available, each containing a single-point mutation from the wild-type sequence. In our analysis, we use the melting temperature as experiment mutation effects, i.e., the response,  while features generated from each sequence, either ESM-based alone or combined with features from CATParc or PSICOV, serve as predictors. The melting temperature is the temperature at which half of the proteins in the sample population are denatured, which measures the protein thermostability. 

\begin{table}[h]
\caption{Spearman correlations between predicted and experimental mutation effects}
\label{table:esm}
\centering
\scalebox{0.9}{
\begin{tabular}{ccccc}
\toprule
                &                & $k$NN                           & SVM   & RF    \\ \midrule
ESM             & Spearman Corr. & 0.793                           & 0.791 & 0.716 \\ \cmidrule{2-5} 
(1280 features) & Std.           & 0.013                           & 0.014 & 0.013 \\ \midrule
ESM+CATParc        & Spearman Corr. & \textbf{0.817} & 0.808 & 0.747 \\ \cmidrule{2-5} 
(1282 features) & Std.           & 0.012                           & 0.013 & 0.016 \\ \midrule
ESM+PSICOV      & Spearman Corr. & 0.758                           & 0.767 & 0.713 \\ \cmidrule{2-5} 
(1282 features) & Std.           & 0.023                           & 0.016 & 0.019 \\ \bottomrule
\end{tabular}
}
\end{table}

Following the procedure in \cite{rives2021biological}, we evaluate the performance of various methods using Spearman correlation between predicted and actual experimental mutation effects. The average Spearman correlations, shown in Table \ref{table:esm}, are computed over $50$ replicates. For each replicate, the data is split into two subsets: $80\%$ as the training set for algorithm training and the remaining $20\%$ as the testing set for calculating the Spearman correlation. Notably, adding features generated by CATParc improves the performance of all algorithm compared to using ESM-based features alone, with the $k$NN emerging as the best-performing method, increasing the Spearman correlation by $3\%$. In contrast, adding features generated from PSICOV reduces the performance of all three algorithms.  Recent studies have shown that large protein language models such as ESM2, despite being trained on massive and diverse sequence databases, yield general-purpose embeddings that may not be fully optimized for specialized downstream tasks, particularly when the task depends on fine-grained structural or evolutionary signals (e.g., \cite{schmirler2024fine}). In our setting, mutational-effect prediction for a specific protein family can benefit from additional features that encode family-specific dependence structure. The features generated by CATParc capture the pairwise dependence structures at the position level derived from the MSA and therefore complement the global biochemical representation provided by ESM embeddings. Although these additional features are low-dimensional, incorporating them consistently improves predictive accuracy in our experiments. This suggests that our proposed CATParc provides task-adapted structural information that is not explicitly represented in pretrained embeddings, highlighting the value of integrating coevolution-based features with modern protein language models. Further details on ESM and the cross-validation procedures are provided in the supplementary materials. These results confirm that integrating our newly developed features with ESM-based features can effectively enhance the prediction of mutation effects.

\section{Conclusions and discussions}
We have developed a novel model-free framework for analyzing protein mutations. By applying one-hot encoding to the MSA data of a specific protein family, we construct a partial correlation graph where each node represents a sequence position as a vector. Contact prediction is reframed as a statistical inference task to determine whether two nodes, corresponding to two positions on a protein sequence, are connected. The proposed spectrum-based test statistic demonstrates superior performance compared to existing $\ell_2$- and $\ell_{\infty}$-norm-based methods, offering higher power and better control of Type I errors. To the best of our knowledge, this is the first approach to reformulate contact prediction and mutation analysis as a statistical inference problem without relying on strong modeling assumptions about the distribution of protein sequences. Furthermore, our framework extends to identifying specific amino acid combinations that contribute to contacts through a valid statistical testing procedure. This capability enables a deeper investigation of residue interactions and enhances our understanding of compensatory mutations.

In this paper, we employ a spectrum-based test statistic to detect individual edges or pairs of positions in a partial correlation graph among multivariate categorical variables.   In the literature on Gaussian graphical models, simultaneous testing procedures for a class of edges of arbitrary size, constructed based on test statistics for individual edges, have been proposed using Gaussian approximation techniques \citep{chang2018confidence}. In principle, our spectrum-based test statistic can also be applied for such tests. However, theoretical analysis requires accurate estimation of a collection of partial correlations to approximate the covariance structure among dependent test statistics for individual edges. We leave this aspect for future research. Additionally, one can also consider multiple testing problems with dependent tests for individual pairs of positions. Since the number of contacts is expected to be small relative to the total number of position pairs, the multiple testing procedure proposed by \citep{cai2016large} can be applied to identify contacts while effectively controlling the false discovery rate. 

\bigskip
\begin{center}
{\large\bf SUPPLEMENTARY MATERIALS}
\end{center}

Technical proofs, extra simulation results, and descriptions of protein families for the contact prediction study are included in the online supplementary materials. Package \texttt{CATParc} on the GitHub repository (\url{https://github.com/FanY4098/CATParc}) documents related codes to reproduce results in this paper.

\bibliographystyle{Chicago} 
\bibliography{arxiv_reference}

\end{document}